\documentclass[preprint2]{aastex}
\shorttitle{Heliosphere as a source of ENAs}
\shortauthors{Czechowski, Grygorczuk, and McComas}

\begin{document}


\title{Heliosphere for a wide range of interstellar magnetic field\\
strengths as a source of energetic neutral atoms}


\author{A. Czechowski\altaffilmark{1} and J. Grygorczuk\altaffilmark{2}}
\affil{Space Research Centre, Polish Academy of Sciences, Bartycka 18A, 
00-716 Warsaw, Poland}

%\author{C. D. Biemesderfer\altaffilmark{4,5}}
%\affil{National Optical Astronomy Observatories, Tucson, AZ 85719}
%\email{aastex-help@aas.org}

\and

\author{D. J. McComas\altaffilmark{3}}
\affil{Southwest Research Institute, San Antonio, TX 78228 USA\\
\and Dept. of Physics and Astronomy, University of Texas at San Antonio,
San Antonio, TX 78249, U.S.A.}


\altaffiltext{1}{e-mail: ace@cbk.waw.pl}
\altaffiltext{2}{e-mail: jolagry@cbk.waw.pl}
\altaffiltext{3}{e-mail: dmccomas@swri.edu}


\begin{abstract}
    
Observations of the energetic neutral atoms (ENAs) of heliospheric origin 
by IBEX differ from expectations based on heliospheric models. It was 
proposed that the structure of the heliosphere may be similar to the 
"two-stream" model derived in 1961 by Parker for the case of strong 
interstellar magnetic field.

Using MHD simulations, we examine possible structure of the heliosphere 
for a wide range of interstellar magnetic field strengths, with different 
choices of interstellar medium and solar wind parameters. For the model 
heliospheres, we calculate the fluxes of ENAs created in the inner 
heliosheath, and compare with IBEX observations.

We find that the plasma flow in the model heliospheres for strong 
interstellar field ($\sim$20 $\mu$G) has a "two-stream" structure, which 
remains visible down to $\sim$5 $\mu$G. The obtained ENA flux 
distribution show the features similar to the "split tail" effect 
observed by IBEX. In our model, the main cause of this effect is the two 
component (fast and slow) solar wind structure.

\end{abstract}


%\keywords{ Magnetohydrodynamics - Sun: heliosphere- ISM: magnetic fields }




\section{Introduction}

Energetic neutral atoms (ENAs) created in the distant heliosphere
by energetic ion neutralization provide a means to remotely observe
the distant regions of the heliosphere. Theoretical models of the 
large scale structure of the heliosphere are important for understanding
and interpreting these observations. 

The global models of the stellar wind interaction with the interstellar 
medium (ISM), leading to the formation of the astrospheres (the heliosphere in 
the case of the Sun) were first introduced in the classic work by  
\citet{parker61}. As shown by Parker, in the case of a star moving 
through unmagnetized interstellar plasma, the stellar wind flow, after 
passing the termination shock, turns ultimately in the direction 
opposite to the motion of the star, forming the "tail" (heliotail). 
This structure was indeed obtained in all models of the heliosphere 
based on numerical solutions of the gas dynamical or MHD equations.

Recently, another class of these models was included in the discussion. 
As again shown by \citet{parker61}, in the case of a star at rest 
with respect to the interstellar medium with strong magnetic field, the 
stellar wind may form, instead of a single astrotail, two oppositely 
directed streams or "drainage plumes", parallel and antiparallel to 
the magnetic field 
direction. Although this model is not strictly applicable to the case 
of the Sun (which is known to move relative to the local interstellar 
medium), a structure of this kind was tentatively discussed in the 
context of the heliospheric observations by IBEX and INCA 
 \citep{mccomas2009,krimigis2009}.

In a recent study, \citet{mccomas2013} followed on the
suggestion of \citet{kivelson2013} and 
made this idea more explicit. \citet{mccomas2013} discovered
that the ENA flux from the 
region of the heliotail in the IBEX data shows two regions shifted to 
opposite sides from the downwind direction (anti-apex of the interstellar 
medium flow), and suggested that the splitting of the heliotail may 
be caused by the effect of a strong interstellar magnetic field
and subsonic interaction \citep{mccomas2012a}. The 
structure of the heliosphere would then be somewhere between the "two stream" 
case (corresponding to the extremely strong magnetic field) and the more 
conventional one-tail structure, corresponding to the weak 
magnetic field. \citet{kivelson2013} suggest also that, in
the case of strong interstellar field, ion heating by reconnection at the 
heliopause may produce a new ENA source.

Among many numerical models of the heliosphere (astrosphere) that take 
the interstellar magnetic field into account, we know of only two published 
models that approach close to the "two stream" extreme field case. 
\citet{florinski2004} report a simulation for the case 
of interstellar flow directed along the magnetic field, but they found 
no stationary solution. For a similar case, \citet{pogorelov2011} found 
a solution with the heliopause open in 
both interstellar upwind and downwind directions. The case of very strong
magnetic field oblique to the interstellar flow was, however, not considered.
As a consequence, there are currently no models of the 
heliosphere that could be used to examine the possibility suggested by 
\citet{mccomas2013}. Such models would also be interesting
in the case of astrospheres, for which 
different combinations of the magnetic field strength and the velocity 
of the star relative to local interstellar medium may be encountered.

In this work we apply a 3D MHD code to obtain global models of the 
heliosphere for a wide range of the interstellar field strengths, from
2 $\mu$G to 20 $\mu$G. For the latter value, our models include the 
analogue of the two-stream Parker model when the star is at rest. 
We follow the evolution of the two-stream structure when the magnetic
field strength decreases and its dependence on the neutral hydrogen 
background. 

For these models, we calculate the directional distribution of the ENA
flux produced inside the heliosphere, between the termination shock and
the heliopause. Our goal is to show how the global structure of the 
heliosphere at different interstellar field strengths would be reflected 
in the ENA observations. In particular, we find that the plasma stream
directions correspond to peaks in the ENA flux distribution.

For moderate interstellar field strengths (2 - 5 $\mu$G) and two-component
(slow + fast) solar wind, our calculations
produce the ENA flux depletions reminiscent of the "split tail" effect observed
in the IBEX data \citep{mccomas2013,schwadron2014}.

We use the following abbreviations: $B_{IS}$, $V_{IS}$, $n_{IS}$, $n_H$ stand
for the interstellar medium magnetic field strength, plasma speed, proton number
density and neutral hydrogen number density, respectively.
$V_{SW}$  denotes the solar wind speed at the inner boundary 
and $n_{SW, 1 AU}$ the solar wind electron density at 1 AU. 
The ENA flux and plasma mass flux directional distributions are presented as
all-sky maps in solar ecliptic J2000 coordinates using Mollweide projection
with the ISM downwind (anti-apex) direction in the center and the ISM upwind
(apex) direction at both far left and far right.

\section{Models of the heliosphere}

Our models of the heliosphere (Figs. \ref{f20v0}, \ref{f20_5}, 
\ref{fmass}, \ref{f5xyz}) are defined by time-stationary 
numerical solutions of the (one-fluid) MHD equations for solar wind 
plasma expanding into the 
magnetized interstellar medium. The MHD code used in our calculation 
was shown to be successful in 
treating the cases of different interstellar magnetic field strengths 
and orientations \citep{ratkiewicz98} 

We consider a range of values (2 to 20 $\mu$G) of the interstellar magnetic 
field strength $B_{IS}$ (Table \ref{table1}). Most of the remaining 
parameters are based on observations. The direction of ${\bf B_{IS}}$ 
(solar ecliptic longitude 221$^o$, 
latitude 39$^o$) is close to the centre of the IBEX ribbon 
\citep{funsten2009}. The direction and magnitude 
(longitude 79$^o$, latitude -4.98$^o$, $V_{IS}$=23.2 km/s) of the 
velocity of the interstellar flow are as 
given by \citet{mccomas2012a}. 
%The angle between 
%$B_{IS}$ and $V_{IS}$ is then equal to 48.34$^o$ (or, equivalently, 
%131.66$^o$). 
Other interstellar and solar wind parameters ($n_{IS}$, 
$n_H$, $V_{SW}$, $n_{SW,1AU}$) are listed 
in Table \ref{table1}. The solar wind at the inner 
boundary is taken either spherically symmetric or to consist of two 
components: the slow wind within 36$^o$ from the solar equator 
plane, and the fast wind elsewhere. The slow and fast components
we take to have the same ram pressure \citep{lechat2012}. To 
avoid numerical reconnection, 
the solar magnetic field is neglected (set to zero at the inner boundary). 

The neutral gas component is treated as 
a constant background. The value $n_H$ of the order 
0.2 cm$^{-3}$ in the 
interstellar medium is presently favored (\citet{izmodenov2009}; 
\citet{bzowski2009}; \citet{zank2013}). However, our MHD calculations 
assume $n_H$=constant 
everywhere. If this constant is equal to the interstellar value for 
$n_H$, this overestimates the hydrogen density in the inner heliosphere, 
leading in particular to underestimation of the termination shock 
distance from the Sun. For this reason in most of our calculations we 
choose $n_H$=0.1 cm$^{-3}$, in agreement with estimations of $n_H$ at 
the termination shock \citep{bzowski2009} rather 
than 0.2 cm$^{-3}$.

Numerical calculations are done on a
spherical ($r$,$\theta$,$\phi$) grid with logarithmic spacing for $r$ and
equal spacing for the angles. The directions of the undisturbed interstellar 
field (${\bf B}_{IS}$) and of the inflow velocity of the interstellar matter 
(${\bf V}_{IS}$) together with the position of the Sun define the (x,y) plane.  
The angle $\theta$ is counted from the apex
direction of the interstellar matter inflow (-x axis) and the  angle $\phi$
from the y axis in the (y,z) plane. The numbers of grid points 
($n_r$,$n_\theta$,$n_\phi$) equal (348,90,180) for most calculations. 
For comparison, some calculations were done on smaller grids. 
The calculational domain lies between the inner boundary at $r$=15 AU 
(in some cases, 30 AU) and the outer boundary at $r$=4500 AU.

\section{Models of the energetic ion distribution and 
calculation of the ENA flux}

We are interested in energetic neutral hydrogen atoms 
in the IBEX energy range ($\sim$0.7 to $\sim$4.3 keV) coming 
from the inner heliosheath (between the termination shock and 
the heliopause). 

The main production mechanism is neutralization
of the parent ions (energetic protons) by picking electrons from 
low energy neutral atoms entering the heliosphere from the 
interstellar medium. The ENAs from the IBEX ribbon (created presumably 
outside the heliosphere) are not considered. 

Since Voyager 2 observations imply that the
bulk plasma temperature downstream of the termination shock is 
low ($\sim$15 eV), we assume that most of the parent ions of 
the ENA derive from the pick-up protons created in the solar wind 
and further accelerated near the termination shock.

Our MHD simulations do not provide the pick-up ion distribution,
so that it must be calculated separately. 

First, inside the termination shock we replace the constant background 
model of the neutral hydrogen distribution used in the code by the 
distribution obtained by using a "hot model" approach 
\citep{thomas78} . We take into account the ionization losses, but 
assume that the gravity is compensated by the radiation pressure.
   
The calculated neutral hydrogen distribution is used to derive
the density of the pick-up protons arriving at the termination shock.
An example of the result is shown in Fig. \ref{shpui}. The density is
nonuniform, decreasing by a factor of $\sim$2 towards the ISM anti-apex
direction.

The distribution function of the energetic protons at the termination
shock we describe by an analytical model (see the Appendix). The model
satisfies the requirement (following from Voyager 2 observations) that 
most (0.8) of the solar wind energy 
upstream of the shock is transferred to energetic particles instead of 
heating the bulk plasma. Figure \ref{fspecsh} shows the example of
the resulting 
spectrum on the opposite sides of the fast/slow wind boundary. The 
pick-up ion density in the fast wind region is lower by a factor of 
$\sim$4-5 than in the slow wind.

The energetic proton flux in the region between the shock 
and the heliopause is then obtained by assuming that the particles are 
convected from the shock along plasma streamlines.  
Losses by neutralization due to charge exchange with ambient neutral hydrogen 
are taken into account. Also included is the effect of adiabatic 
acceleration: the energy of the energetic particle (in the plasma frame)
varies along a streamline.

The ENA flux $J_{ENA}$ arriving in the inner solar system we 
calculate by integrating the ENA production rate along the lines of sight 
corresponding to the grid directions: 
\begin{equation}
J_{ENA}=\int ds J_{ion}\sigma_{cx}n_H
\label{eqjena}
\end{equation}
where the integral is over the distance $s$ along the line of sight 
(between the termination
shock and the heliopause), $J_{ENA}$ and $J_{ion}$ the fluxes of the
ENA and of the parent ions, respectively (both at the same energy and
directed along the line of sight), $\sigma_{cx}$ the charge-exchange 
cross section at the ENA energy (we neglect the speed of the neutral H
atoms from the background), and $n_H$ the number density of the
neutral H background. 

The characteristic distance for the neutralization loss for the energetic
protons convected from the termination shock is given by $V/\beta_{cx}$,
where $V$ is the plasma speed in the inner heliosheath 
and $\beta_{cx}$ the rate for charge-exchange between the energetic 
proton and the low energy hydrogen atom. In the IBEX energy range,
$\beta_{cx}$=0.7-1 10$^{-8}$ s$^{-1}$. In the region of the heliosheath not
far from the termination shock, $V$$\sim$75 km/s 
for our models with $V_{SW}$=400 km/s, and $V$$\sim$120 km/s for the models with 
$B_{IS}$=20 $\mu$G which assume $V_{SW}$=750 km/s. The value of $V/\beta_{cx}$
is therefore 50-70 AU (400 km/s solar wind) or 80-110 AU (750 km/h solar wind).
Most of the production of the ENA from the termination shock accelerated protons
must therefore take place within this distance. Note, however, that the fast
solar wind effects extend this range by increasing $V$.  

The losses of the ENA on the way to the observation 
point are not included in the present calculations, but are small for
the higher IBEX energies (\citet{bzowski2008}; 
\citet{mccomas2012b})

\section{Results: Structure of the model heliospheres}

We concentrate on the aspects of the heliospheric structure which are 
most important for understanding the production of the ENA in the inner
heliosheath: (1) the termination shock and the energetic ion distribution
in the vicinity of the shock, (2) the structure of the plasma flow downstream from
the shock (Figures \ref{f20v0}, \ref{f20_5},
\ref{f5xyz}), which determines the transport of the energetic ions from the
termination shock region into the 
inner heliosheath, and (3) the shape and size of the heliopause.  

In Table \ref{table1} we collected some results concerning the geometry of the 
heliosphere for our models: minimum (r$_{TS,min}$) and maximum (r$_{TS,max}$) 
distances to the termination shock, minimum distance (r$_{HP,min}$) to the 
heliopause, the maximum height ("height") of the heliosphere counted in the $z$
direction (perpendicular to the ($B_{IS}$,$V_{IS}$) plane), and the width 
("width") of the heliosphere along the straight line passing through the point 
of maximum height and perpendicular to the ISM inflow direction.

\subsection{Plasma flow structure}

\begin{figure}
\centering
%\epsscale=0.4
%\plotone{figs/hptrajyz.020sym0.04_vis0.1.eps}
\includegraphics[width=8cm]{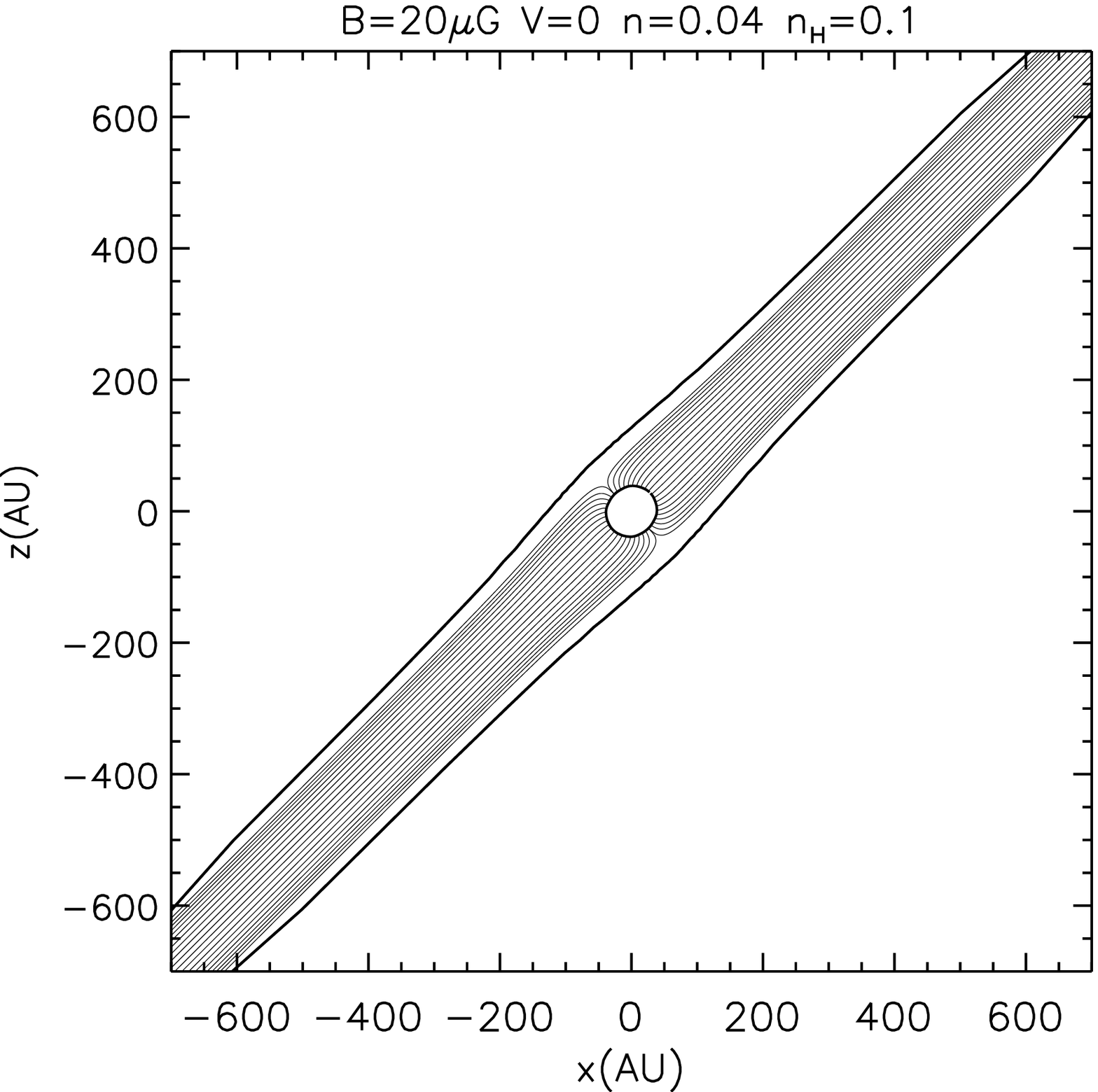}
\caption{Flow lines of the solar plasma for the MHD solution
corresponding to the hypothetical heliosphere with the Sun at rest 
($V_{IS}$=0) relative to the 
interstellar medium with plasma density $n_{IS}$=0.04 cm$^{-3}$, neutral
hydrogen density $n_{H}$=0.1 cm$^{-3}$ and very strong interstellar 
magnetic field ($B_{IS}$=20 $\mu$G). The solar wind is spherically 
symmetric with $V_{SW}$=750 km/s, $n_{SW, 1 AU}$=4.2 cm$^{-3}$. 
The solution is similar to
the Parker model \citep{parker61}. The heliopause 
and the termination shock are shown by thick lines.
\label{f20v0}}
\end{figure}

\begin{figure}
\centering
%\epsscale=0.4
%\plotone{figs/alfstrtraj.020sym0.04_nh0.01c.eps}
%\plotone{figs/alfstrtraj.020sym0.06c.eps}
\includegraphics[width=8.0cm]{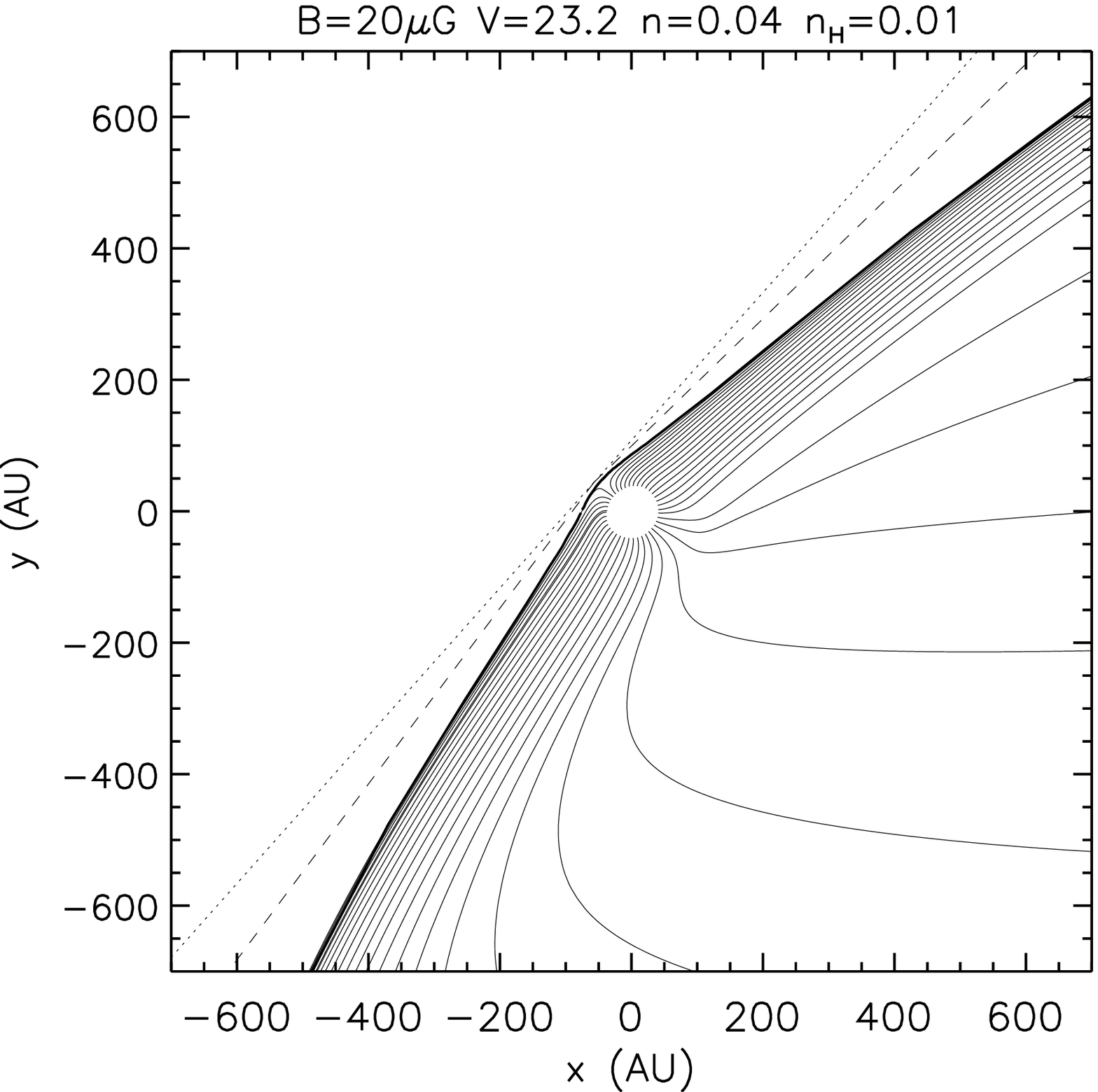}
\includegraphics[width=8.0cm]{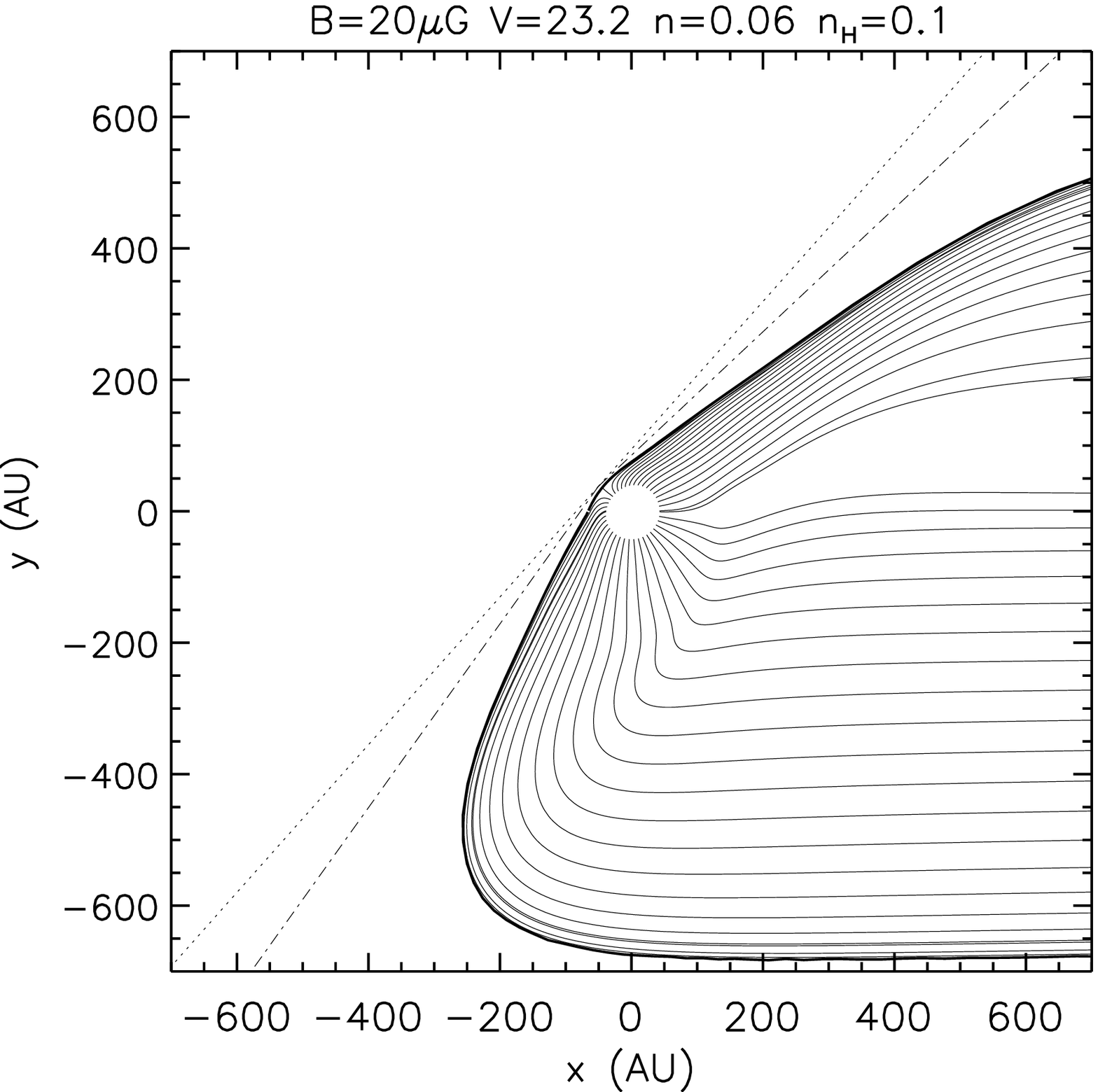}
\caption{Plasma flow lines in the 
(x,y)$\equiv$($\vec{B}_{IS}$,$\vec{V}_{IS}$) plane obtained
from MHD simulations for $B_{IS}$=20 $\mu$G,
$V_{IS}$=23.2 km/s and the solar wind
with $V_{SW}$=750 km/s, $n_{SW, 1AU}$=4.2 cm$^{-3}$, 
The first figure corresponds to $n_H$=0.01 cm$^{-3}$
and $n_{IS}$=0.04 cm$^{-3}$, and the second to  
$n_H$=0.1 cm$^{-3}$
and $n_{IS}$=0.06 cm$^{-3}$. The solar wind is spherically
symmetric,
so that the (x,y) plane is the symmetry plane.
The heliopause is shown by the thick line. The dotted line shows the
$\vec{B}_{IS}$ direction. The dashed lines show 
the Alfven wings (see Eq. \ref{eqalfv}).
\label{f20_5}}
%\label{f20_5}
\end{figure}

\begin{figure}
%\centering
%\epsscale{0.6}
%\plotone{figs/flu_20sf_5maja.eps}
%\plotone{figs/flu_5sf_5maja.eps}
%\plotone{figs/flu_3sf_5maja.eps}
\includegraphics[width=8.0cm]{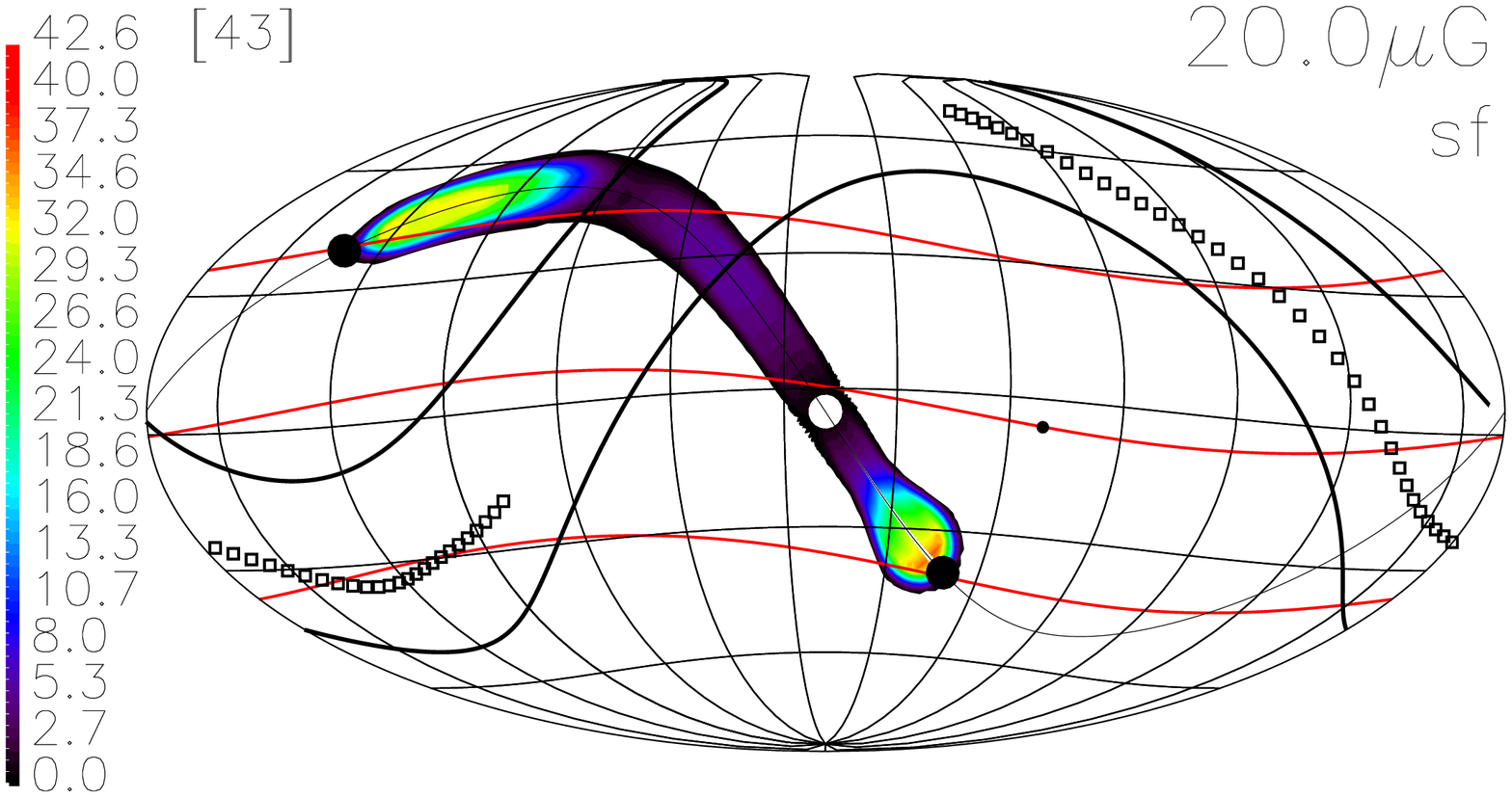}
\includegraphics[width=8.0cm]{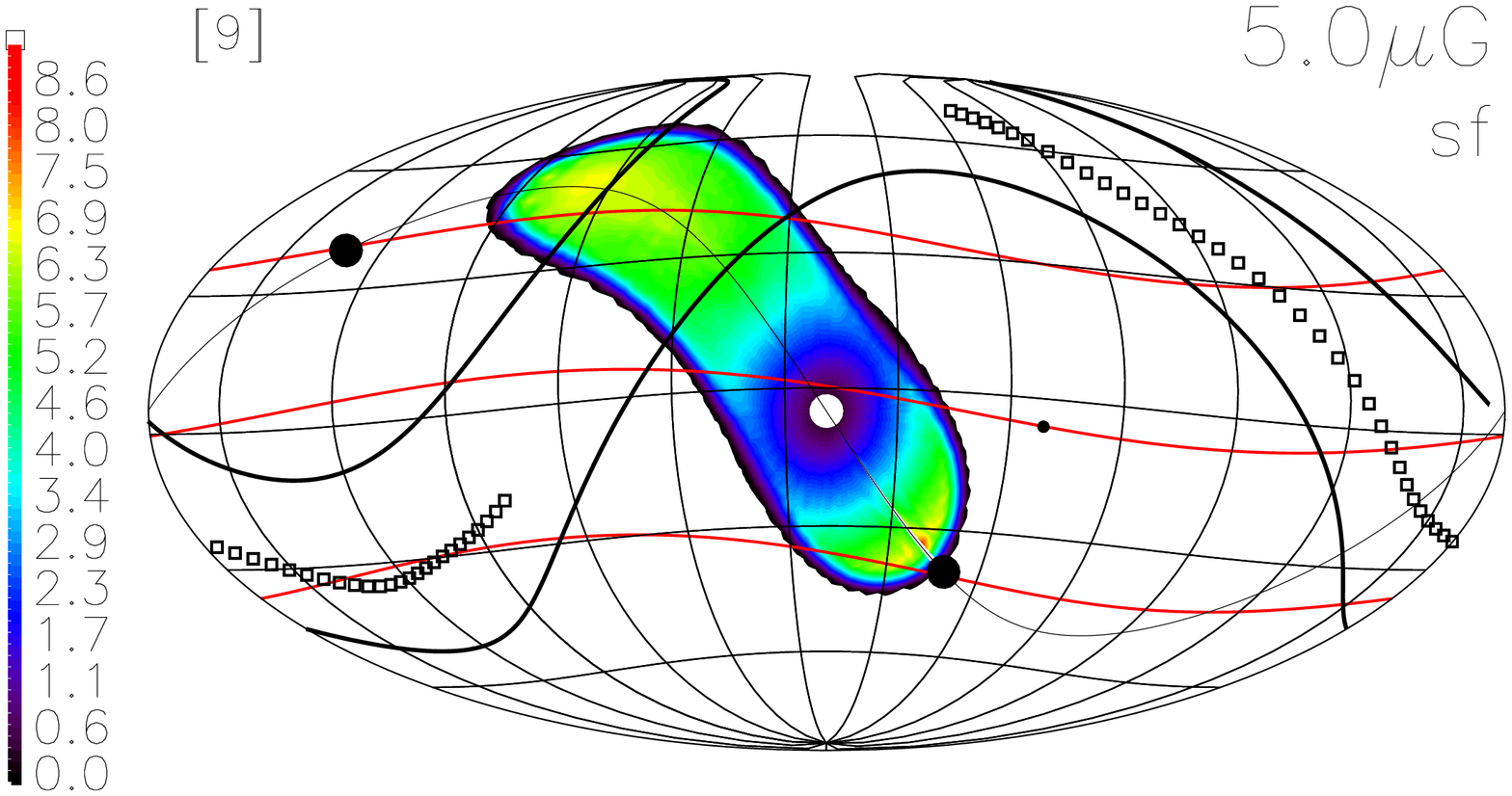}
\includegraphics[width=8.0cm]{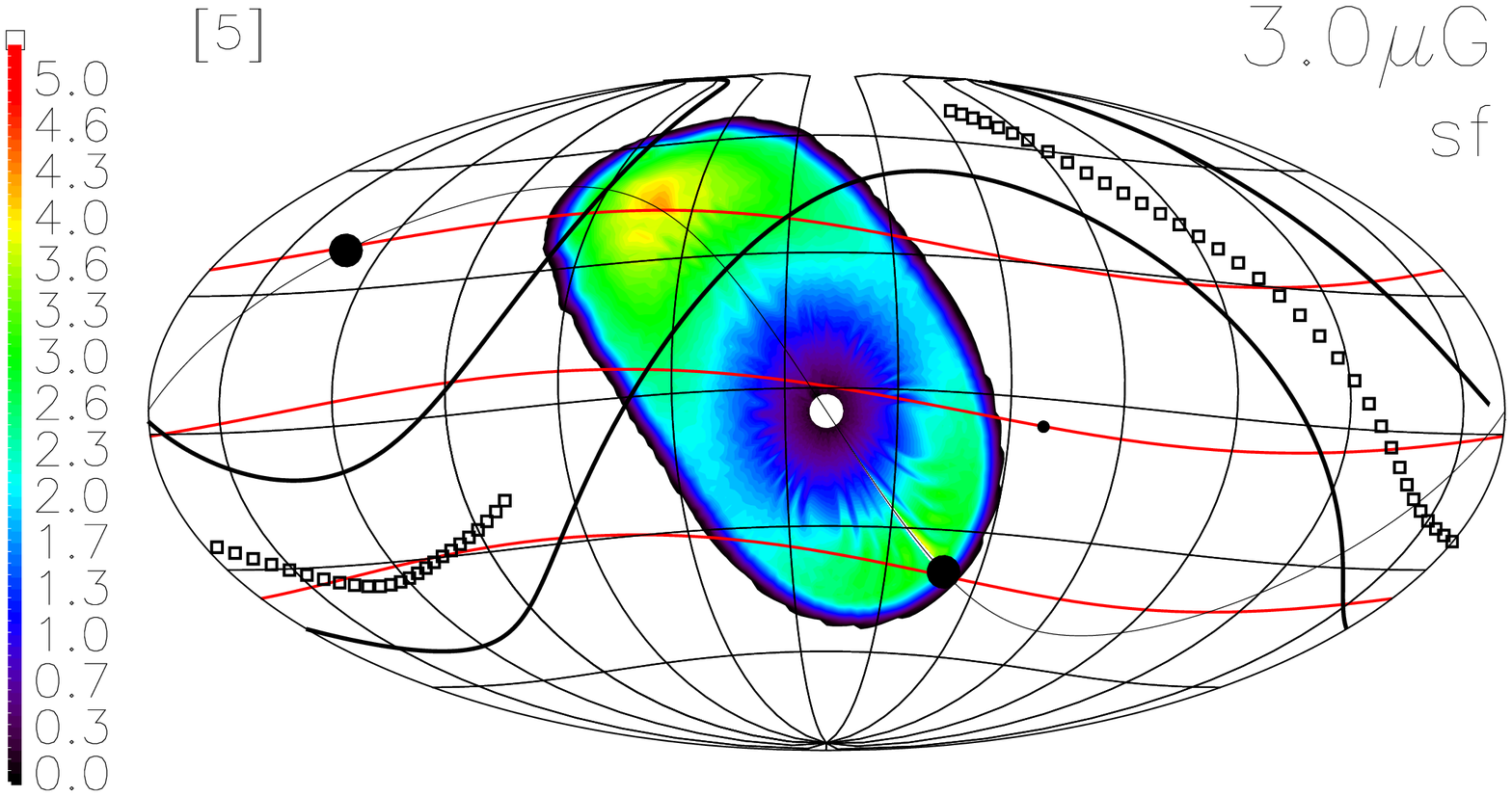}
\caption{Directional distribution of the solar plasma mass flux 
(obtained from streamline density) in units 10$^4$ $m_p$ cm$^{-2}$ s$^{-1}$ 
at a distance 300 AU from the Sun for the cases of $B_{IS}$=20 $\mu$G
(the first panel) 5 $\mu$G and 3 $\mu$G. The values of $V_{IS}$=23.2 km/s, 
$n_H$=0.1 cm$^{-3}$, $n_{IS}$=0.06 cm$^{-3}$ are the same for all cases. 
The solar wind is spherically symmetric, with $V_{SW}$=750 km/s, 
$n_{SW, 1 AU}$=4.2 cm$^{-3}$ (upper panel) and $V_{SW}$=400 km/s, 
$n_{SW, 1 AU}$=5.55 cm$^{-3}$ for the remaining cases.
The projection on the celestial sphere
is the same as used for the ENA distributions (Figs. \ref{fenaasym}, 
\ref{fenan2} and \ref{fenasym}). It is centered on the anti-apex direction 
of the ISM flow (white circle). The interstellar field and anti-field 
directions are marked by black circles.\label{fmass}}
%\label{fmass}
\end{figure}

\begin{figure}
\centering
%\epsscale=0.4
%\plotone{figs/velo_5asym_nh0.1_bw_max76.eps}
%\plotone{figs/velo_5asym_nh0.1_xz.eps}
\includegraphics[width=8cm]{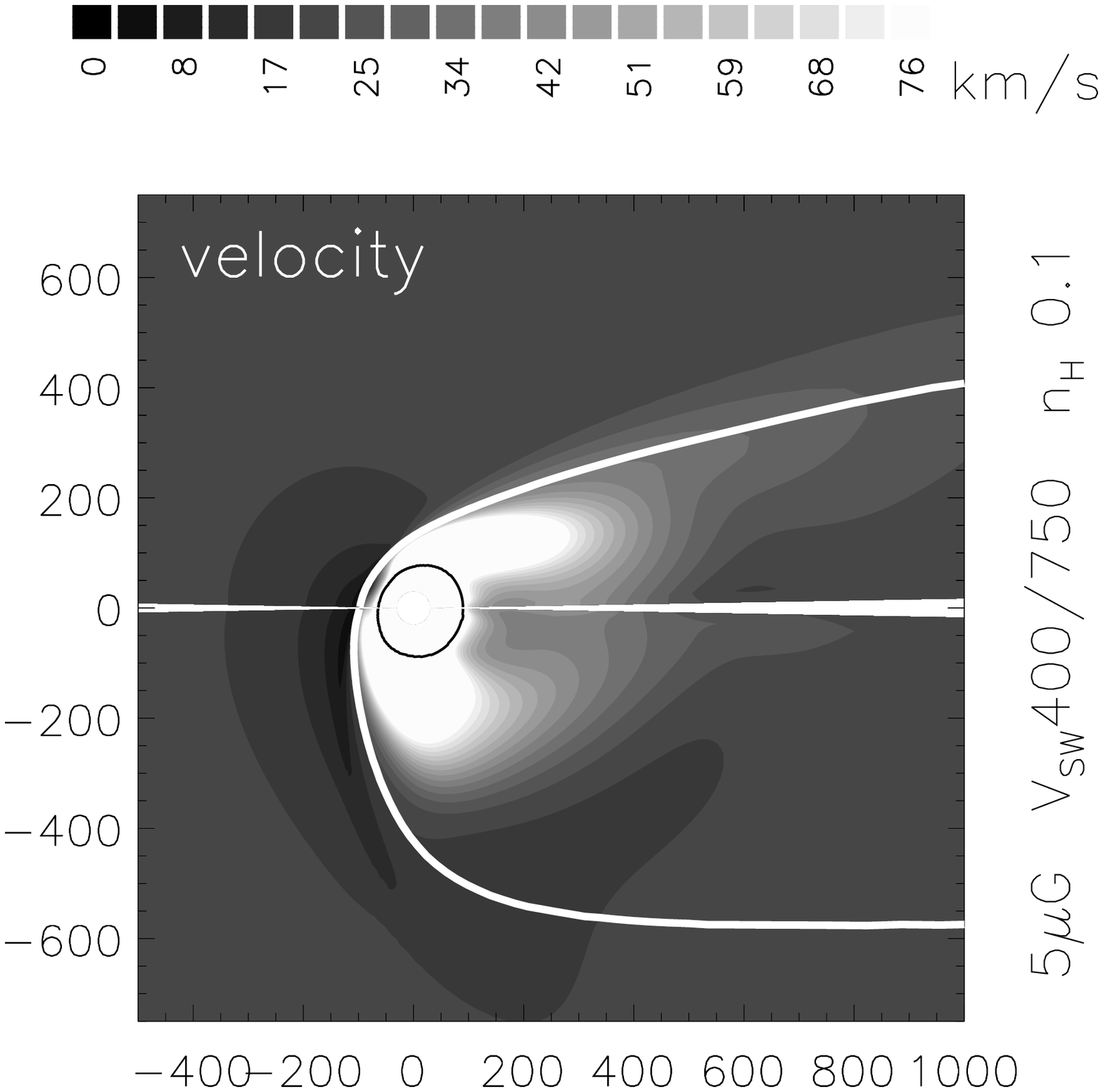}
\includegraphics[width=8cm]{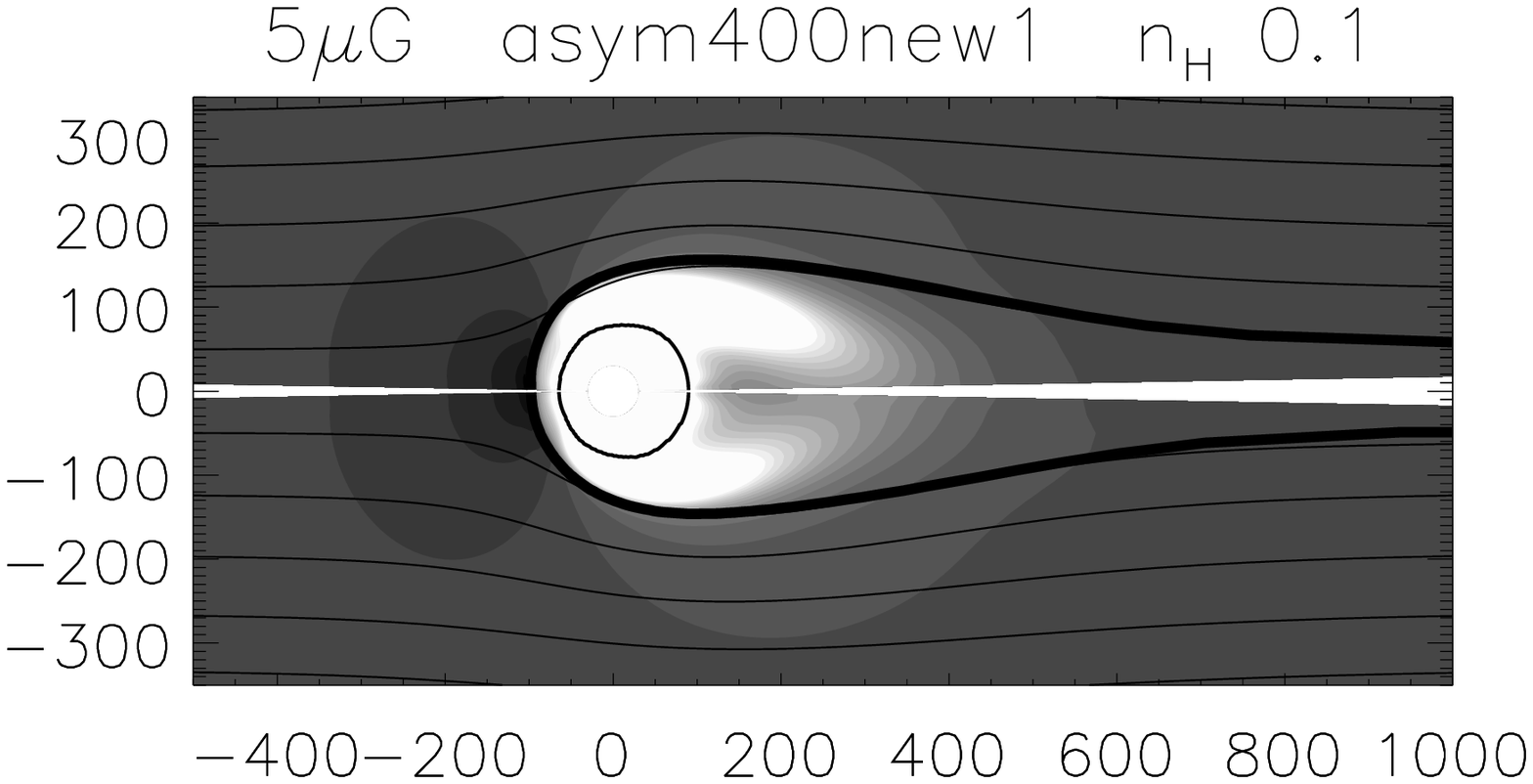}
\caption{Plasma velocity distributions in the (x,y) plane (first panel)
and the (x,z) plane (second panel) for 
$B_{IS}$=5 $\mu$G, $n_H$=0.1 cm$^{-3}$, $V_{IS}$=23.2 km/s, 
$n_{IS}$=0.06 cm$^{-3}$. The solar wind is asymmetric (fast/slow) with
$V_{SW}$=750/400 km/s, $n_{SW, 1AU}$=1.58/5.55 cm$^{-3}$.
The outlines of the heliopause and of the termination shock are also shown.
Note the flattening of the heliosphere by the asymmetric pressure of the
interstellar magnetic field.
\label{f5xyz}}
%\label{f5xyz}
\end{figure}

\begin{figure}
\centering
%\epsscale=0.4
%\plotone{figs/ploshpui2_5sym.eps}
%\plotone{figs/ploshpui2_5asym.eps}
\includegraphics[width=9cm]{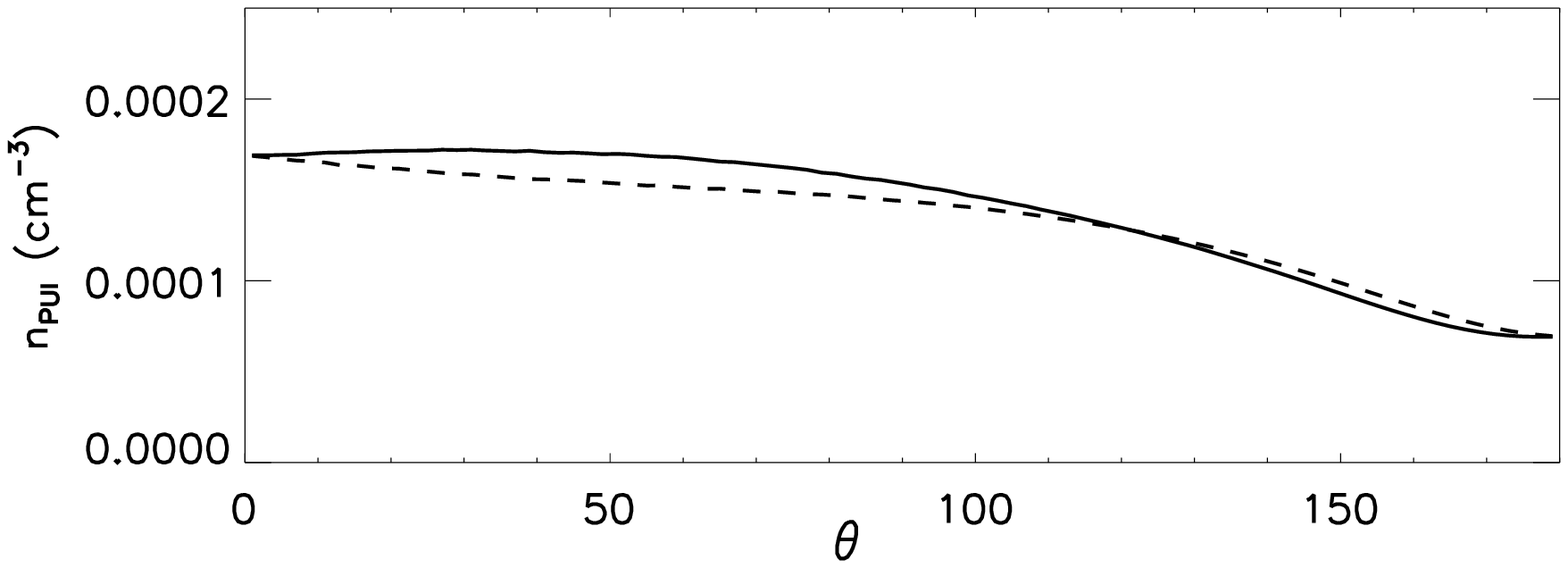}
\includegraphics[width=9cm]{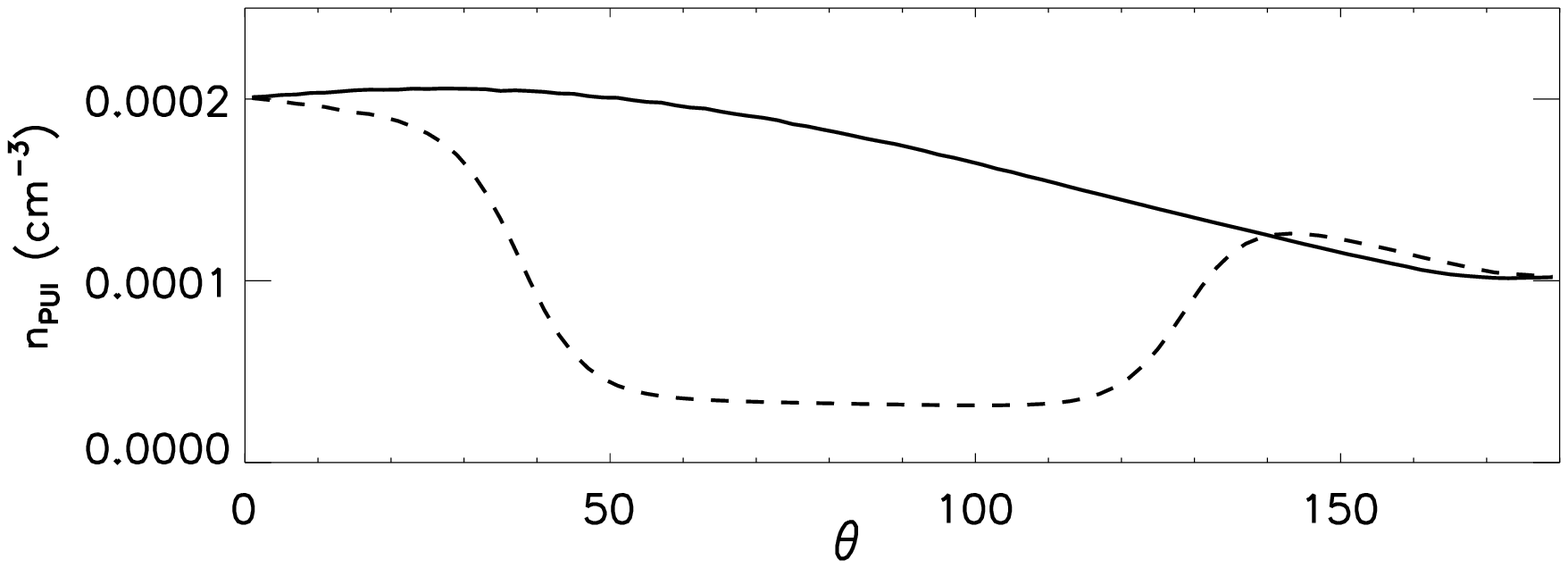}
\caption{The pick-up proton density at the termination shock as a function
of angle $\theta$ (counted from the inflow direction of the interstellar
medium) for two choices of the angle $\phi$ corresponding to maximum (solid
line) and minimum (dashed line) densities for intermediate $\theta$.
The cases shown are: $B_{IS}$=5 $\mu$G, symmetric solar wind 
($V_{SW}$=400 km/s, $n_{SW, 1 AU}$=5.55 cm$^{-3}$, the upper panel) and
$B_{IS}$=5 $\mu$G, asymmetric (fast/slow) solar wind
($V_{SW}$=750/400 km/s, $n_{SW, 1 AU}$=1.58/5.55 cm$^{-3}$, lower panel).
The low pick-up proton density region in the lower figure (dashed line) 
corresponds to the fast wind.
The distributions have a similar form for other values of $B_{IS}$, provided
that $V_{IS}$ is the same (23.2 km/s).\label{shpui}}
%\label{shpui}
\end{figure}

\begin{figure}
\centering
%\epsscale=0.4
%\plotone{figs/plospecsh2.005asym0.06.eps}
\includegraphics[width=9cm]{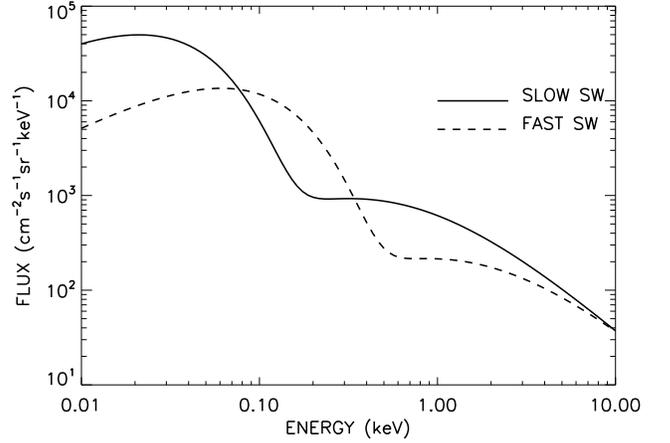}
\caption{The proton flux as a function of energy at two selected locations 
on the termination shock, corresponding to slow (solid line) and fast (dashed line)
solar wind, respectively. The case of $B_{IS}$=5 $\mu$G, asymmetric (fast/slow)
solar wind ($V_{SW}$=750/400 km/s, $n_{SW, 1 AU}$=1.58/5.55 cm$^{-3}$ is 
illustrated.  
The flux of high energy protons up to $\sim$10 keV
is higher in the slow solar wind region because of high slow solar wind density.  
\label{fspecsh}}
%\label{fspecsh}
\end{figure}

For strong $B_{IS}$, the plasma flow between the termination shock and the 
heliopause has a two-stream structure. 
Figures \ref{f20v0} and \ref{f20_5}
show the solar plasma streamlines for $B_{IS}$=20 $\mu$G starting at the 
termination shock in the 
symmetry plane of the solution (for spherically symmetric solar wind).   
For $V_{IS}$=0 (Figure \ref{f20v0}), the streams are respectively parallel 
and antiparallel to $\vec{B}_{IS}$ 
as in Parker's model \citep{parker61}.  
When $V_{IS}\neq$0 (Figure \ref{f20_5}) the streams
appear as bunches of almost parallel streamlines
and include only a part of the flow. The streams are then 
deflected from the $\pm${\bf B}$_{IS}$ directions towards the direction of
{\bf V}$_{IS}$ and run approximately parallel to the 
"wings" of the heliopause (see next subsection).  

One point to note is the effect of neutral hydrogen background on the plasma
streams. For $n_H$=0.1 cm$^{-3}$ (close to the observed value), the momentum 
exchange between the stellar plasma and background hydrogen caused by 
charge-exchange interaction deflects and diffuses the streams (Fig. \ref{f20_5},
second panel). For $n_H$=0.01 cm$^{-3}$ and smaller, the two streams are more 
prominent (Fig. \ref{f20_5}, first panel) and would be recognizable even for 
the weaker field ($B_{IS}$=5 $\mu$G).  

A quantitative representation of the two stream structure can be obtained
by plotting the density of plasma streamlines. Figure \ref{fmass} shows the 
directional distribution (projected onto the celestial sphere) of the density 
of solar plasma streamlines crossing the Sun-centered sphere of radius 
300 AU for three of our models with spherically symmetric solar wind and 
different values of $B_{IS}$. The streamlines start at the termination shock, 
with their initial points chosen to have the same directional distribution as 
the plasma mass flow. The streams appear as two separate density peaks near to 
(but shifted from) the directions parallel 
and antiparallel to the interstellar field. The two-stream structure is most 
prominent at strong $B_{IS}$ (20 $\mu$G) but persists also for $B_{IS}$=5 $\mu$G 
and 3 $\mu$G.

\subsection{Shape of the heliopause}

Asymmetric pressure of the interstellar magnetic field causes the heliosphere to 
expand in the ($B_{IS}$, $V_{IS}$) plane and contract in the perpendicular 
direction (Figure \ref{f5xyz}).
This effect is well known from many numerical simulations.

For strong $B_{IS}$ (20 $\mu$G and, 
if $n_H$ is small, even for 5 $\mu$G) we find that the forward part of the 
heliopause in the ($B_{IS}$, $V_{IS}$) plane has the form of straight 
"wings" (Fig. \ref{f20_5}). 
A similar shape was predicted using the Newtonian approximation 
(\citet{fahr1988}, \citet{czechowski1998}). In 
particular, the angle $\alpha$ between 
the "wings" and the x axis (the $V_{IS}$ direction) was derived: 
\begin{equation} 
\tan\alpha=\frac{\tan\gamma}{1\pm V_{IS}/V_A \cos\gamma},
\label{eqalfv} 
\end{equation} 
where $\gamma$ is the angle between the interstellar magnetic 
field and the x axis, and $V_A$ is the Alfv\'en speed in the interstellar medium.
Equation \ref{eqalfv} is also known from the theory of Alfv\'en wings, which appear 
when a conducting body (like a satellite or a planetary magnetosphere) is moving 
through a magnetized plasma (\citet{drell1965};  
\citet{neubauer1980}; \citet{kivelson2013}; \citet{saur2013}). 

In Fig. \ref{f20_5}, the directions of the Alfv\'en wings are shown by 
dashed-dotted lines. Although our results confirm the presence of straight 
"wings", the angle between each wing and the x axis is not in agreement with 
Eq. \ref{eqalfv}. Note that the interstellar plasma flow
in our simulation is subsonic, so that the use of Newtonian approximation
is not strictly justified.

\subsection{Shape of the termination shock}

For $B_{IS}$=20 $\mu$G, and $V_{IS}$=0 (Figure \ref{f20v0}), we find that
the shape of the termination shock is very close to a Sun-centered sphere, in 
agreement with Parker's model \citep{parker61}. The shock shape is
weakly elongated along the magnetic field direction, but the difference between 
the maximum and minimum radius of the shock is only 8\% 
of the maximum.

In the Parker model \citep{parker61}, the spherical shock appeared as a 
consequence of assuming incompressible flow downstream. In our 
calculations, the plasma density downstream is not constant and the 
approximate symmetry of the shock must have a different explanation. A 
detailed study of this problem is beyond the scope of the present work. 

For $V_{IS}\neq$0, we find that the ratio between the 
Sun - termination shock distances in the ISM anti-apex (heliotail) and 
ISM apex (nose) directions goes down when the interstellar field 
strength increases. The shape of the shock is not much affected by the
asymmetry of the solar wind, because of our assumption that the ram 
pressures of the fast and slow solar wind are equal.

\section{Results: Angular distributions of the ENA flux}

The results for directional distribution of the ENA flux 
calculated for the observer in
the inner solar system are shown in Figures \ref{fenasym}, \ref{fenan2} and
\ref{fenaasym}. 
The projections used in our sky maps (Figs. \ref{fmass}, \ref{fenasym}, 
\ref{fenan2}, \ref{fenaasym} and \ref{fgam5}) are the same as 
in Fig. 6 of 
\citet{mccomas2013} and in Fig. 10 of \citet{schwadron2014}, 
that is centered on the interstellar downwind
direction (white dot in the middle). To help with orientation, 
the position of the IBEX 
ribbon is marked with the line of squares, and the interstellar upfield 
({\bf B}$_{IS}$) and downfield ({\bf -B}$_{IS}$) directions are marked by 
dots. 

\subsection{ENA flux distributions for spherically symmetric solar wind.}

\begin{figure}
\centering
%\epsscale{0.492}
%\plotone{figs/mapka_bis.eps}
%\plotone{figs/flux5_bis5sf_uni.eps}
%\plotone{figs/flux5_bis3sf_uni.eps}
%\plotone{figs/flux5_bis5sf_uni_noacc.eps}
%\plotone{figs/flux5_bis3sf_uni_noacc.eps}
\includegraphics[width=6.5cm]{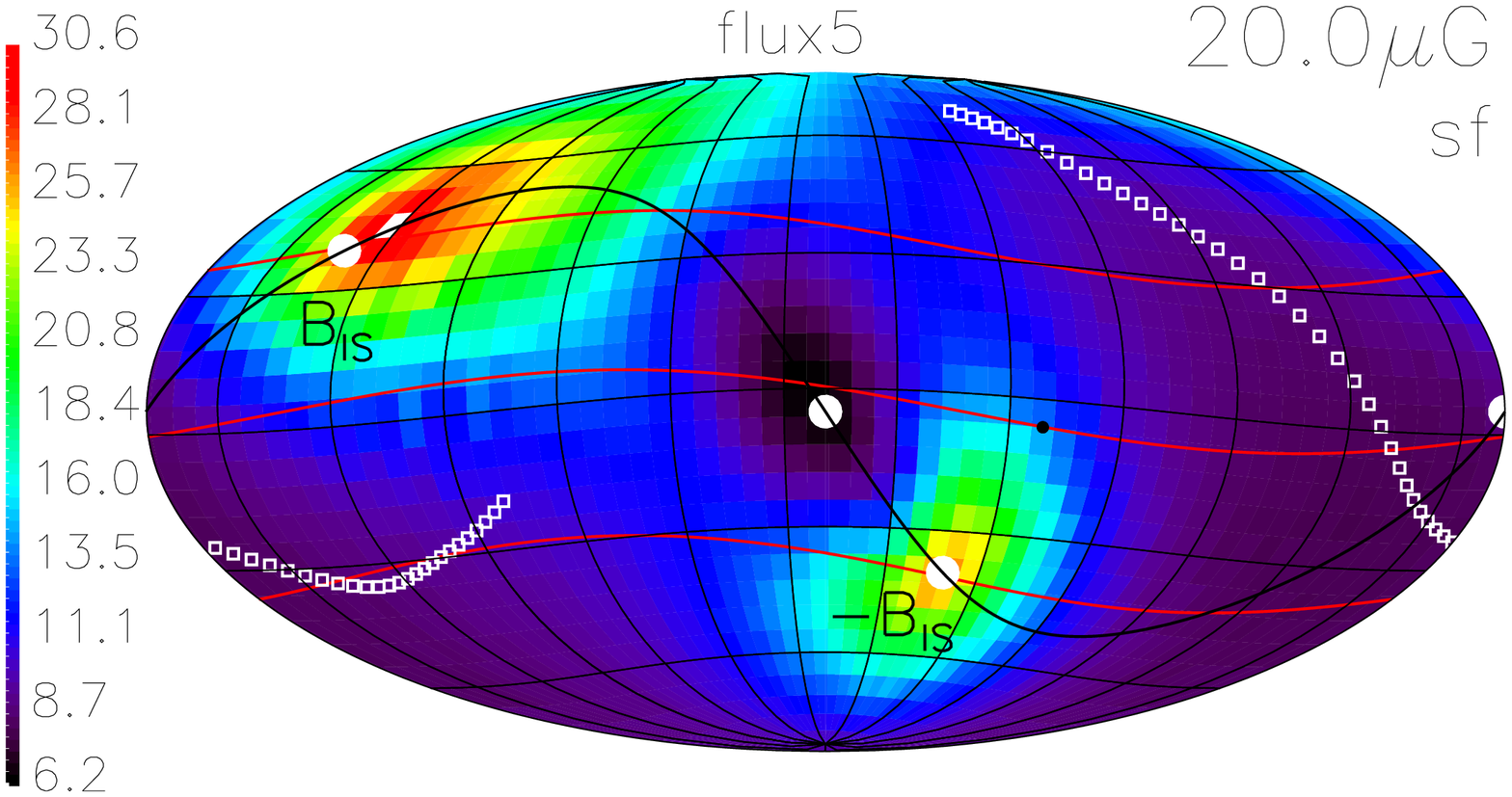}
\includegraphics[width=6.5cm]{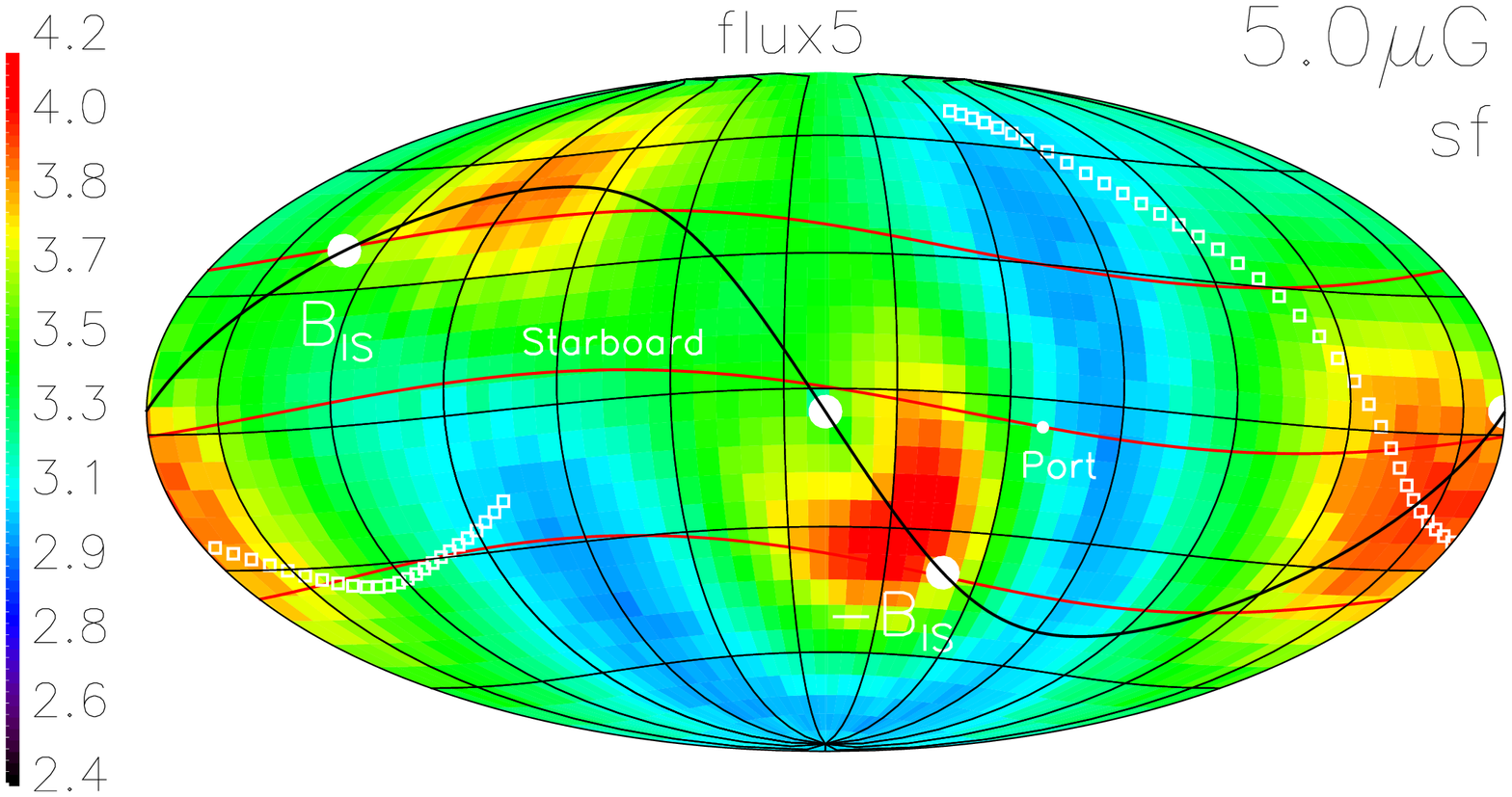}
\includegraphics[width=6.5cm]{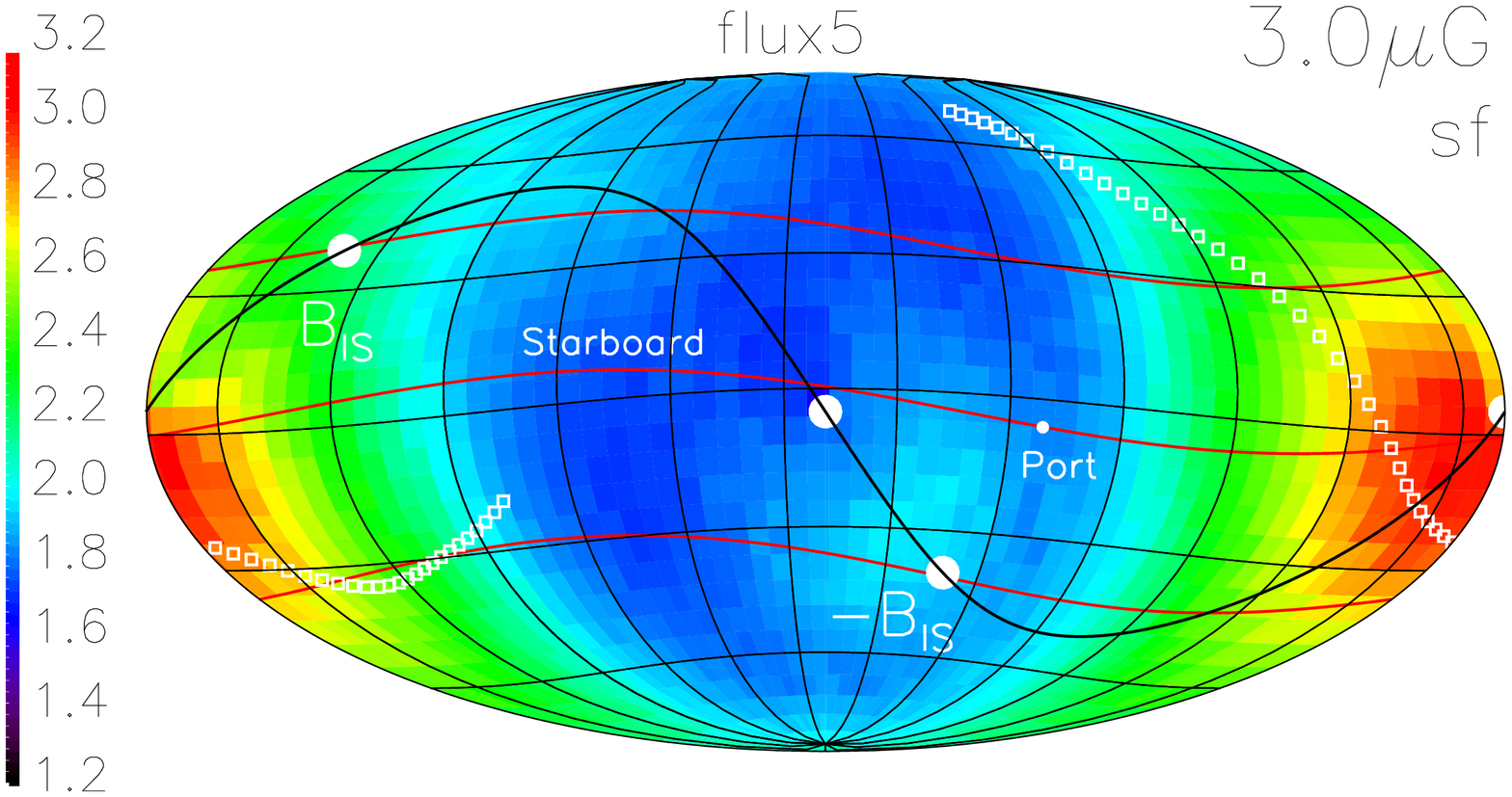}
\includegraphics[width=6.5cm]{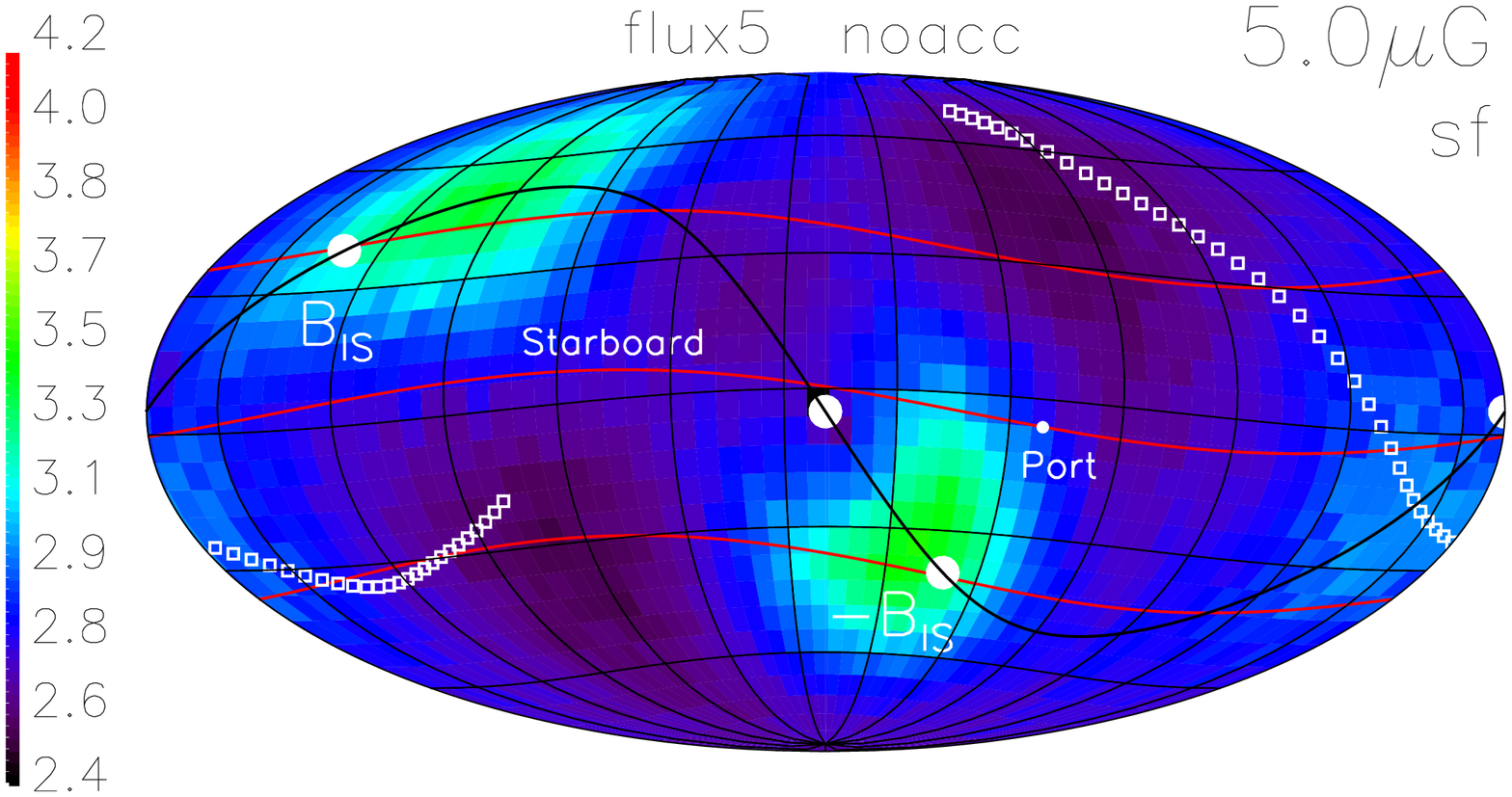}
\includegraphics[width=6.5cm]{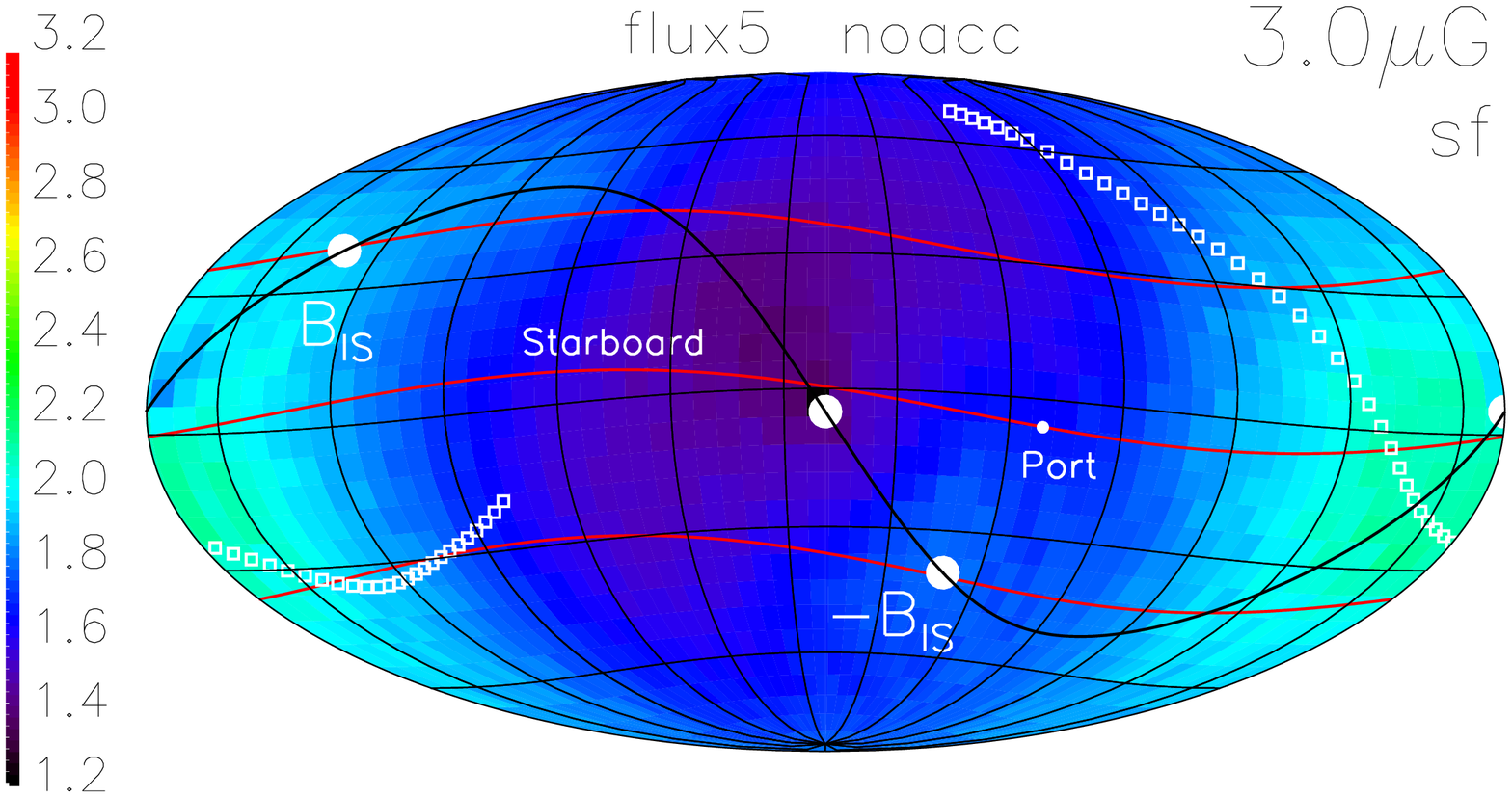}
\caption{ENA flux distribution (4.3 keV) 
in units (cm$^2$ s sr keV)$^{-1}$ for spherically symmetric
solar wind. The cases shown (from left to right and from top to bottom) 
are: $B_{IS}$=20 $\mu$G, 5 $\mu$G, 3 $\mu$G, 5 $\mu$G and 3 $\mu$G,
the two last cases corresponding to adiabatic acceleration switched
off. The ENA flux for $B_{IS}$=20 
$\mu$G and 5 $\mu$G has two peaks corresponding to two streams of the
plasma flow. These streams effectively disappear for 3 $\mu$G. The 
ENA flux from the ISM 
apex region is affected by adiabatic acceleration. 
The values of $V_{IS}$=23.2 km/s, $n_{IS}$=0.06 cm$^{-3}$, 
$n_H$=0.1 cm$^{-3}$, $V_{SW}$=400 km/s, 
$n_{SW, 1AU}$=5.55 cm$^{-3}$ are the same for all figures.
\label{fenasym}}
%\label{fenasym}
\end{figure}

Figure \ref{fenasym} shows our results for directional distributions 
of the 4.3 keV ENA flux for $B_{IS}$=20 $\mu$G, 5 $\mu$G, and 3 $\mu$G
assuming spherically symmetric solar wind.
The distributions are symmetric with 
respect to the ({\bf B}$_{IS}$, {\bf V}$_{IS}$) plane, which is shown as 
a thick black line.
For the case of $B_{IS}$=20 $\mu$G and $B_{IS}$=5 $\mu$G, the ENA flux has 
two peaks at directions approximately parallel and antiparallel to the 
interstellar field, corresponding to the two streams of the solar plasma.
For $B_{IS}$=3 $\mu$G, these peaks almost completely disappear. 

\begin{figure}
\centering
%\epsscale=0.4
%\plotone{figs/lostrlst_020sym0.06wrg.eps}
\includegraphics[width=8.0cm]{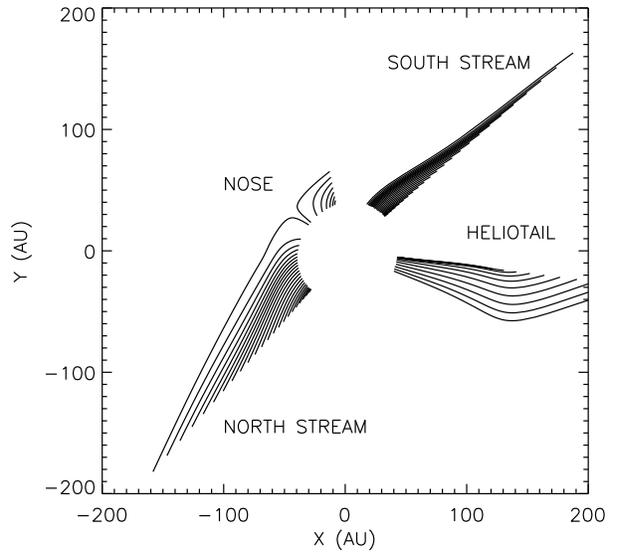}
\caption{Streamlines ending at selected lines-of-sight within
250 AU from the Sun. Two of the lines-of-sight correspond to
peaks of the ENA flux associated with two plasma streams.
In addition, one line-of-sight is in the forward part of the 
heliosphere (the "nose" region)
and one is close to the ISM anti-apex direction (heliotail).
The case illustrated 
corresponds to the 1st panel in Fig. \ref{fenasym}
($B_{IS}$=20 $\mu$G, symmetric solar wind).   
\label{fslin}}
%\label{fslin}
\end{figure}

\begin{figure}
\centering
%\epsscale=0.4
%\plotone{figs/sena5sym_noac.eps}
\includegraphics[width=8.0cm]{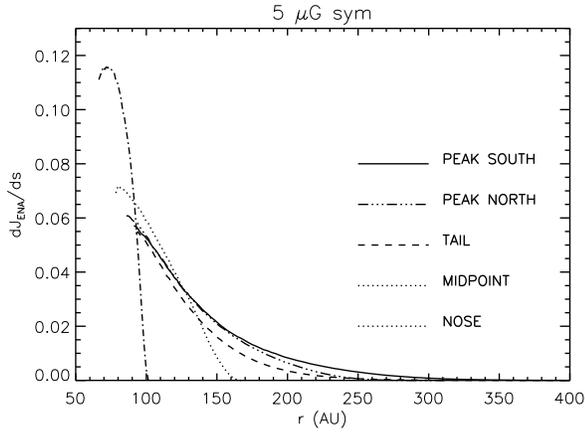}
\caption{Profiles of the ENA production rate (4.3 keV)
along 5 different directions corresponding to: ENA flux
maximum in the south hemisphere; ENA flux maximum in the 
north hemisphere; vicinity of the ISM anti-apex; the middlepoint
between the two maxima of the flux. The case illustrated 
corresponds to the 4th panel in Fig. \ref{fenasym}
($B_{IS}$=5 $\mu$G, symmetric solar wind, adiabatic acceleration
switched off).   
\label{fsymp}}
%\label{fsymp}
\end{figure}

The reason why (for $B_{IS}$=20 and 5$\mu$G) the two solar plasma 
streams are associated with high ENA flux intensity can be explained 
as follows (see Figs. \ref{fslin}, \ref{fsymp}). The lines-of-sight
directed along the streams are close to parallel to the local
plasma flow (Fig. \ref{fslin}), so that Eq. \ref{enastr} is applicable. Since 
$\beta_{cx}L/V\gg1$ in the stream regions, it follows that 
$J_{ENA}\approx(V/v)J_0$. This relation agrees well with our results for
the case of $B_{IS}$=5 $\mu$G without adiabatic acceleration. 

The same argument can be applied to the lines-of-sight in the heliotail.
We find that the ENA flux from these directions is lower than for
the streams because $V$ and $J_0$ are lower for the case of the
heliotail. 

For $B_{IS}$=5 and 3 $\mu$G, a broad maximum of the ENA flux appears in 
the "nose" (ISM apex) region (Fig. \ref{fenasym}, panel 2 and 3).
This is caused by the adiabatic acceleration of the energetic protons 
downstream from the shock and disappears if the adiabatic acceleration 
is switched off (Fig. \ref{fenasym}, panel 4 and 5).

Eq. \ref{enanose} provides the estimation for the ENA flux from the "nose"
direction in the absence of adiabatic acceleration. For $B_{IS}$=5 $\mu$G, 
the parameter $\beta_{cx}L/V_0$$\sim$1/2,
implying that $J_{ENA}\approx(1/3)(V_0/v)J_0$. This agrees well with our
numerical result when the adiabatic acceleration is disregarded 
($J_0$=110 cm$^{-2}$s$^{-1}$sr$^{-1}$keV$^{-1}$, 
$J_{ENA}$=3.8 cm$^{-2}$s$^{-1}$sr$^{-1}$keV$^{-1}$).

\subsection{ENA flux distributions for asymmetric solar wind:
The "split tail" for high energy ENA flux}

\begin{figure}
\centering
%\epsscale=0.4
%\plotone{figs/flux2_bis5asym_uni.eps}
%\plotone{figs/flux2_bis3asym_uni.eps}
\includegraphics[width=7cm]{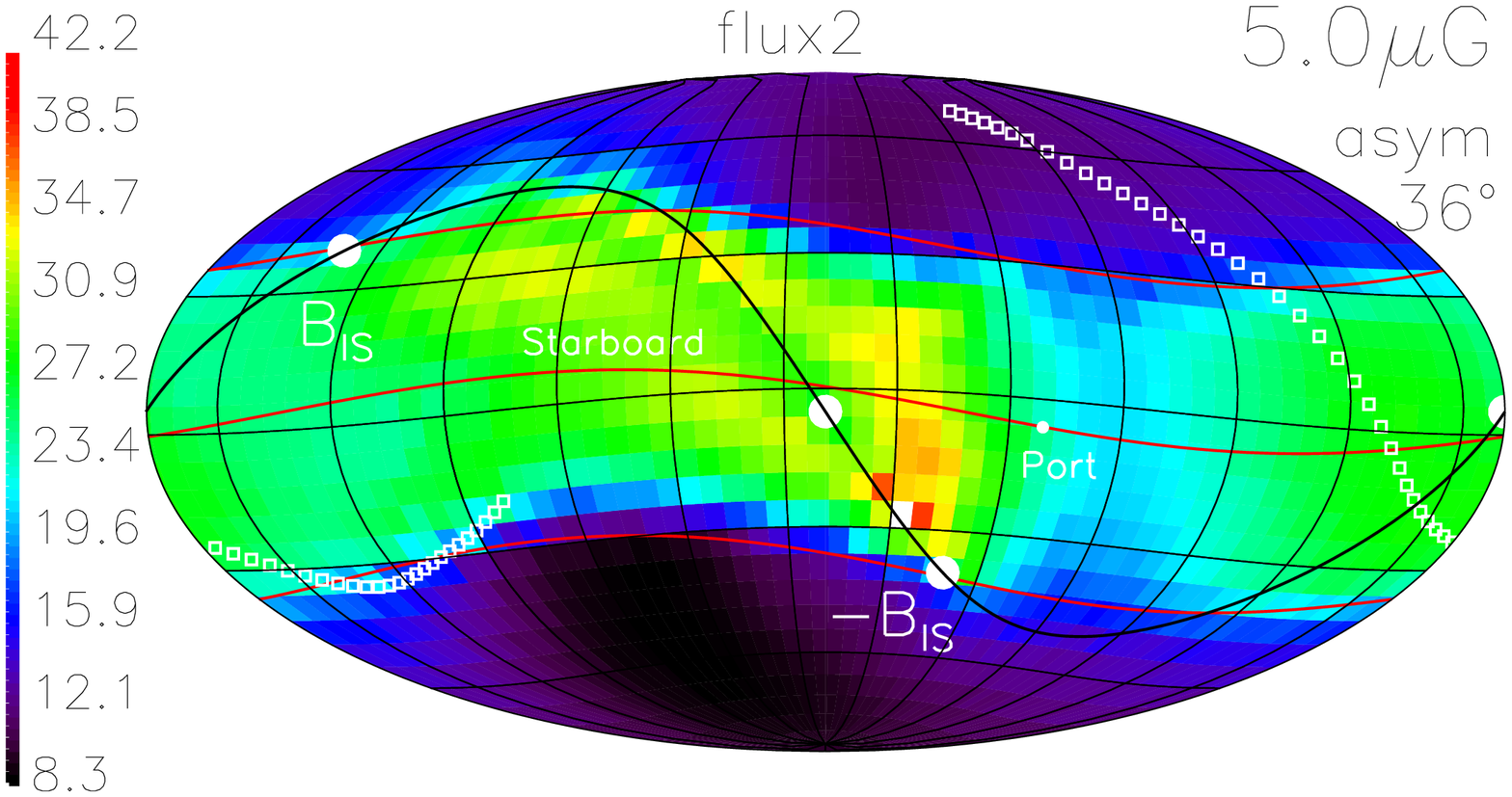}
\includegraphics[width=7cm]{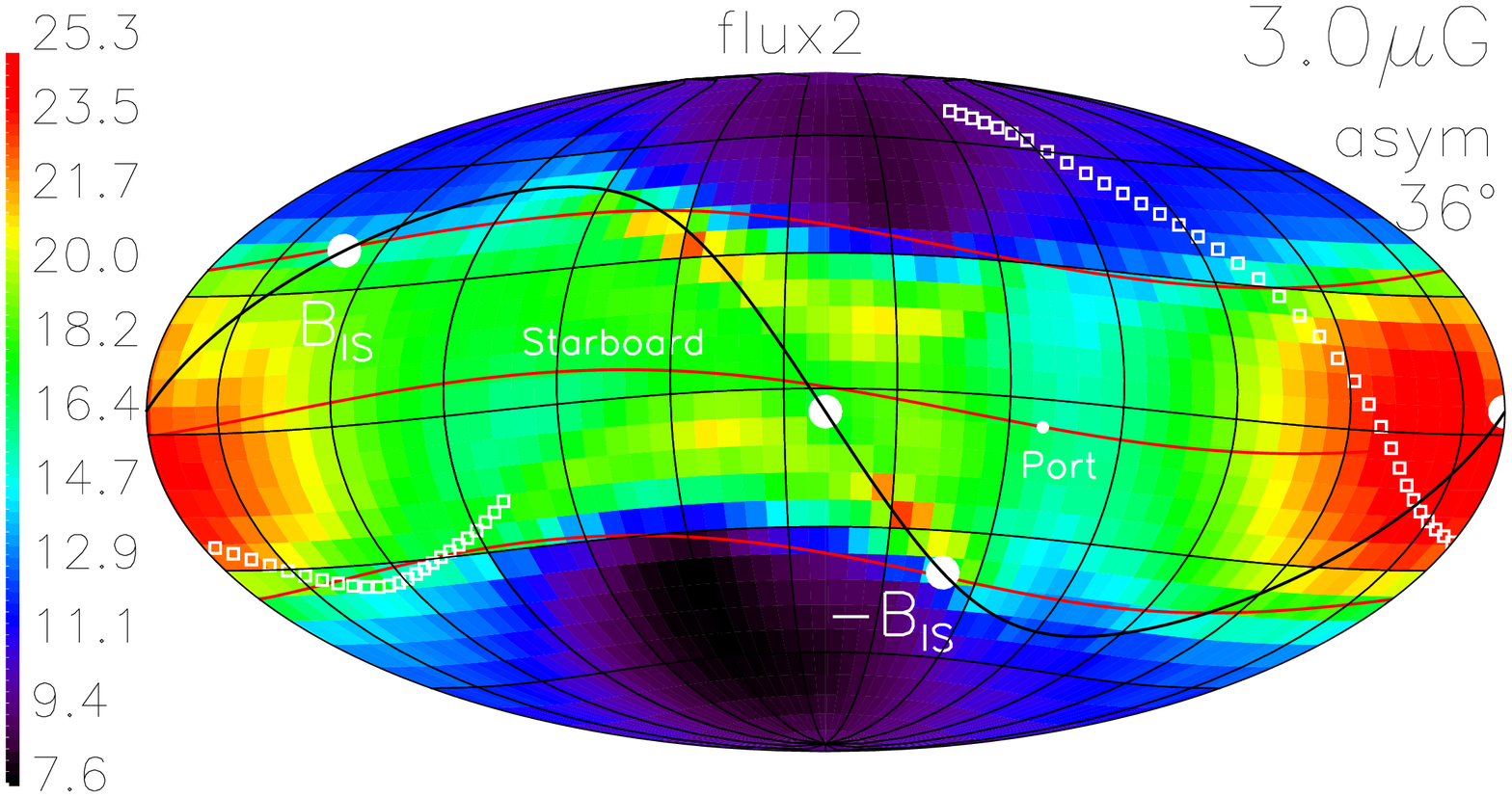}
\caption{ENA flux distribution (1.1 keV) 
in units (cm$^2$ s sr keV)$^{-1}$ for $B_{IS}$=5 $\mu$G
(first panel) and $B_{IS}$=3 $\mu$G (second panel),
for the case of asymmetric (slow+fast)
solar wind, with slow wind contained within $\pm$36$^o$ from 
the equator.
The values of $V_{IS}$=23.2 km/s, $n_{IS}$=0.06 cm$^{-3}$, 
$n_H$=0.1 cm$^{-3}$, $V_{SW}$=750/400 km/s (fast/slow), 
$n_{SW, 1AU}$=1.58/5.55 cm$^{-3}$ (fast/slow) are the same for both cases.
\label{fenan2}}
%\label{fenan2}
\end{figure}

\begin{figure}
\centering
%\epsscale=0.4
%\plotone{figs/flux5_bis15asym_uni.eps}
\includegraphics[width=6.7cm]{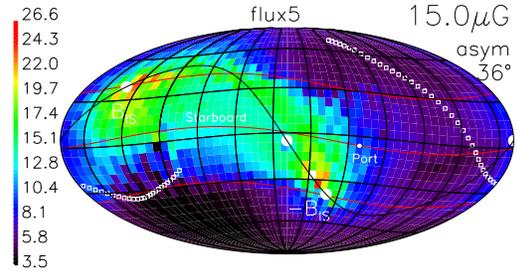}
\caption{ENA flux distribution (4.3 keV) 
in units (cm$^2$ s sr keV)$^{-1}$ for  
$B_{IS}$=15 $\mu$G.
\label{fenaasym15}}
%\label{fenaasym15}
\end{figure}

\begin{figure}
\centering
%\epsscale=0.4
%\plotone{figs/flux5_bis5asym_uni.eps}
%\plotone{figs/flux5_bis3asym_uni.eps}
%\plotone{figs/flux5_bis5asym_uni_noacc.eps}
%\plotone{figs/flux5_bis3asym_uni_noacc.eps}
\includegraphics[width=6.7cm]{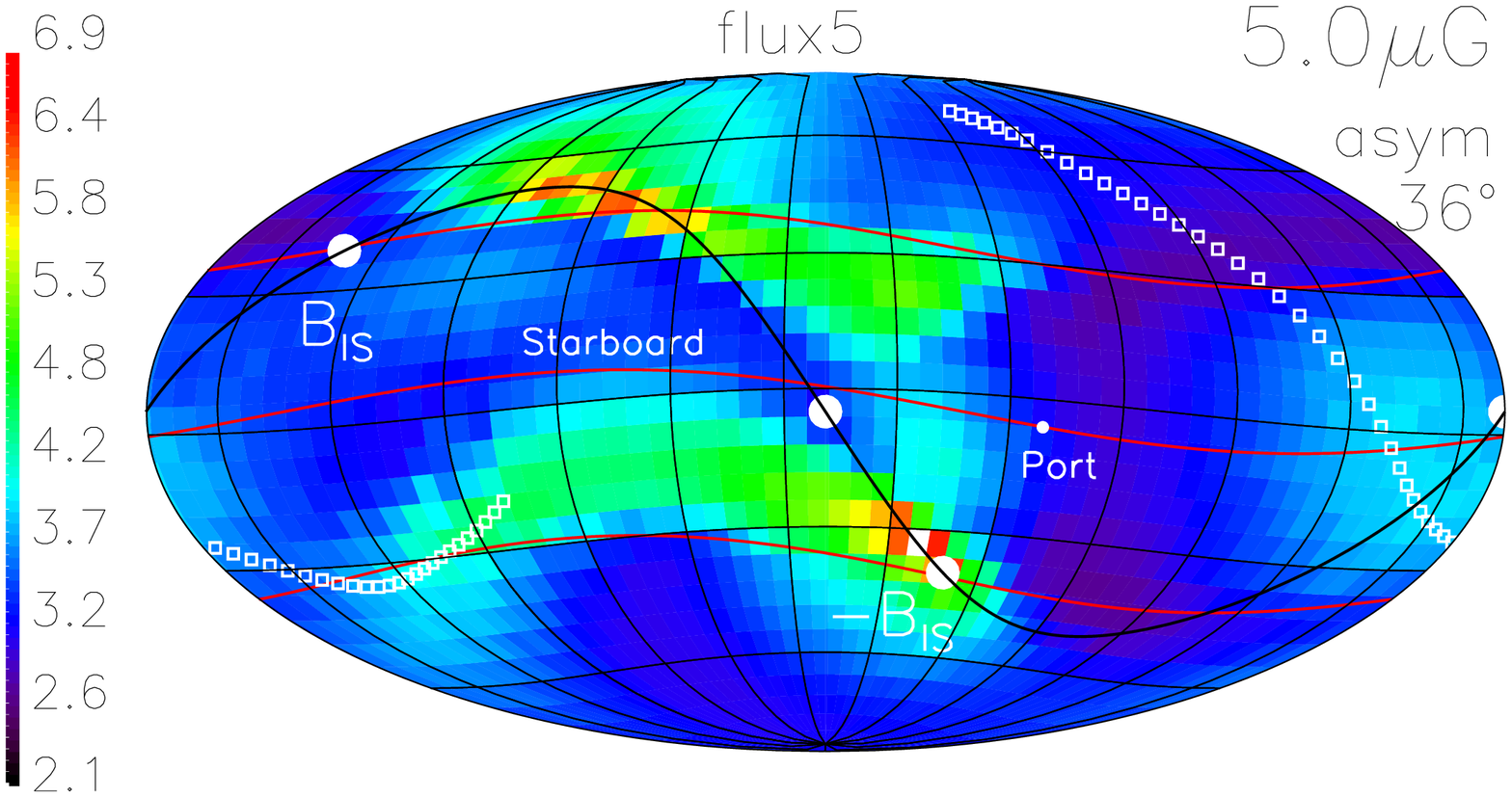}
\includegraphics[width=6.7cm]{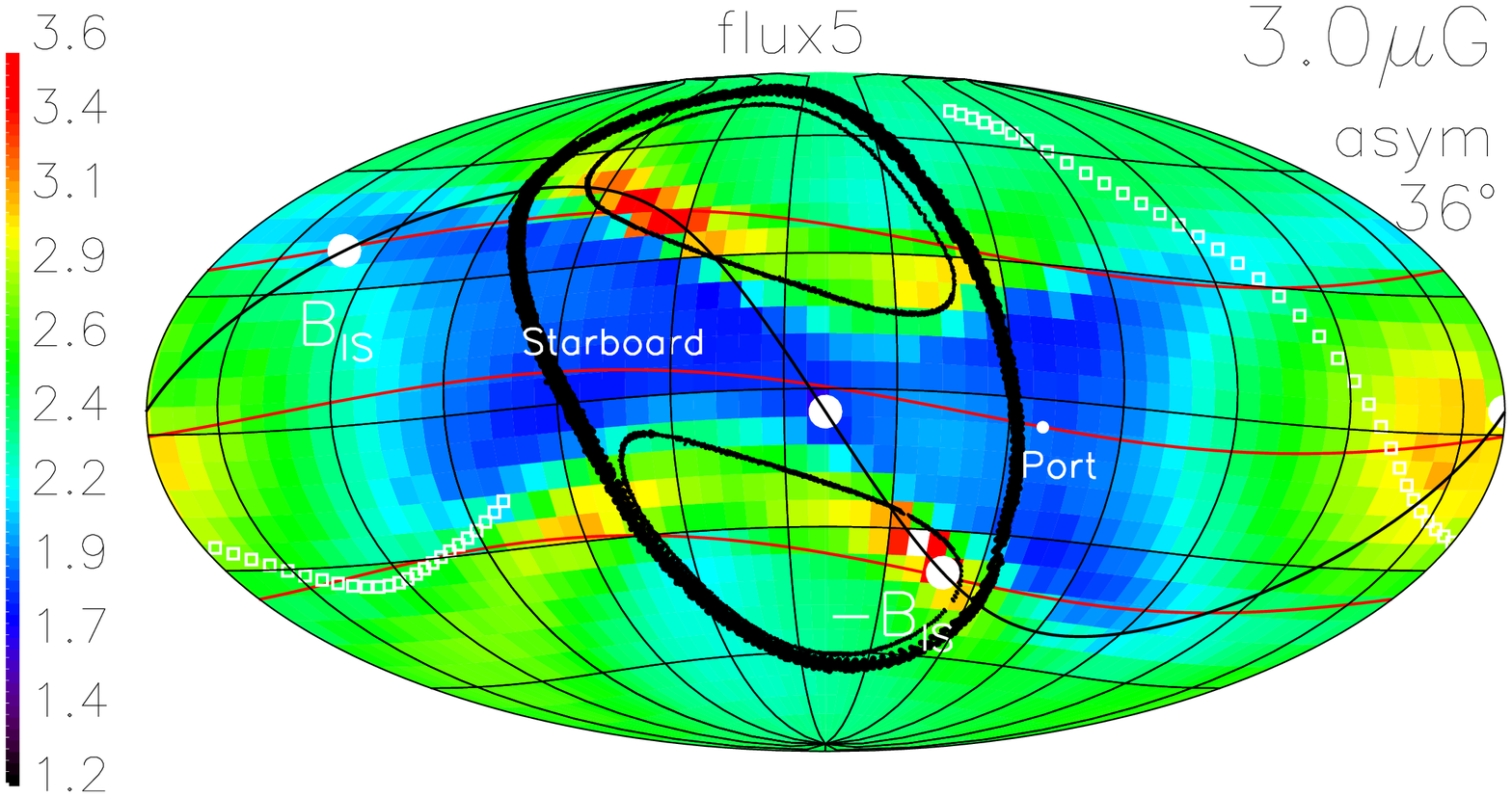}
\includegraphics[width=6.7cm]{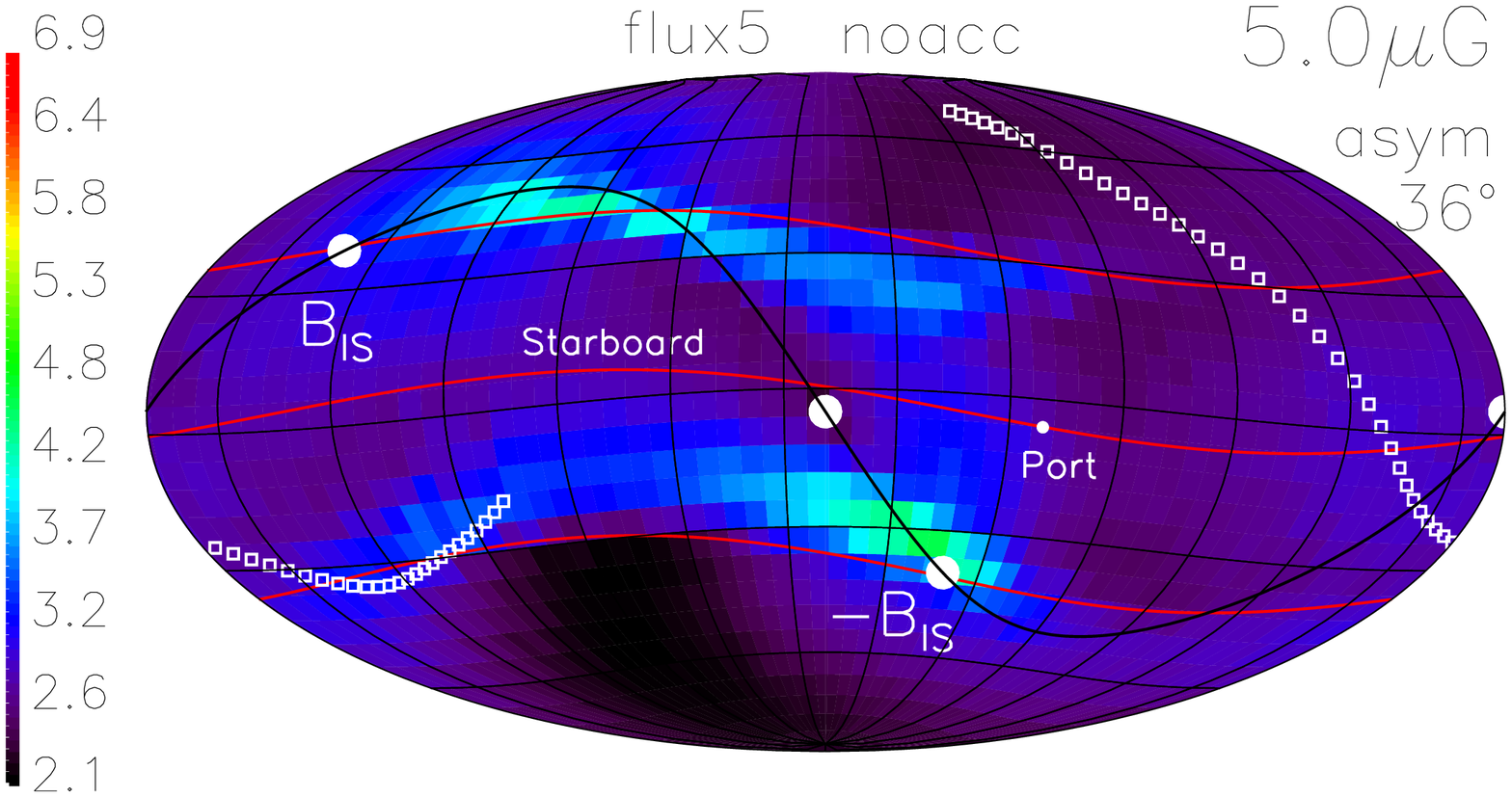}
\includegraphics[width=6.7cm]{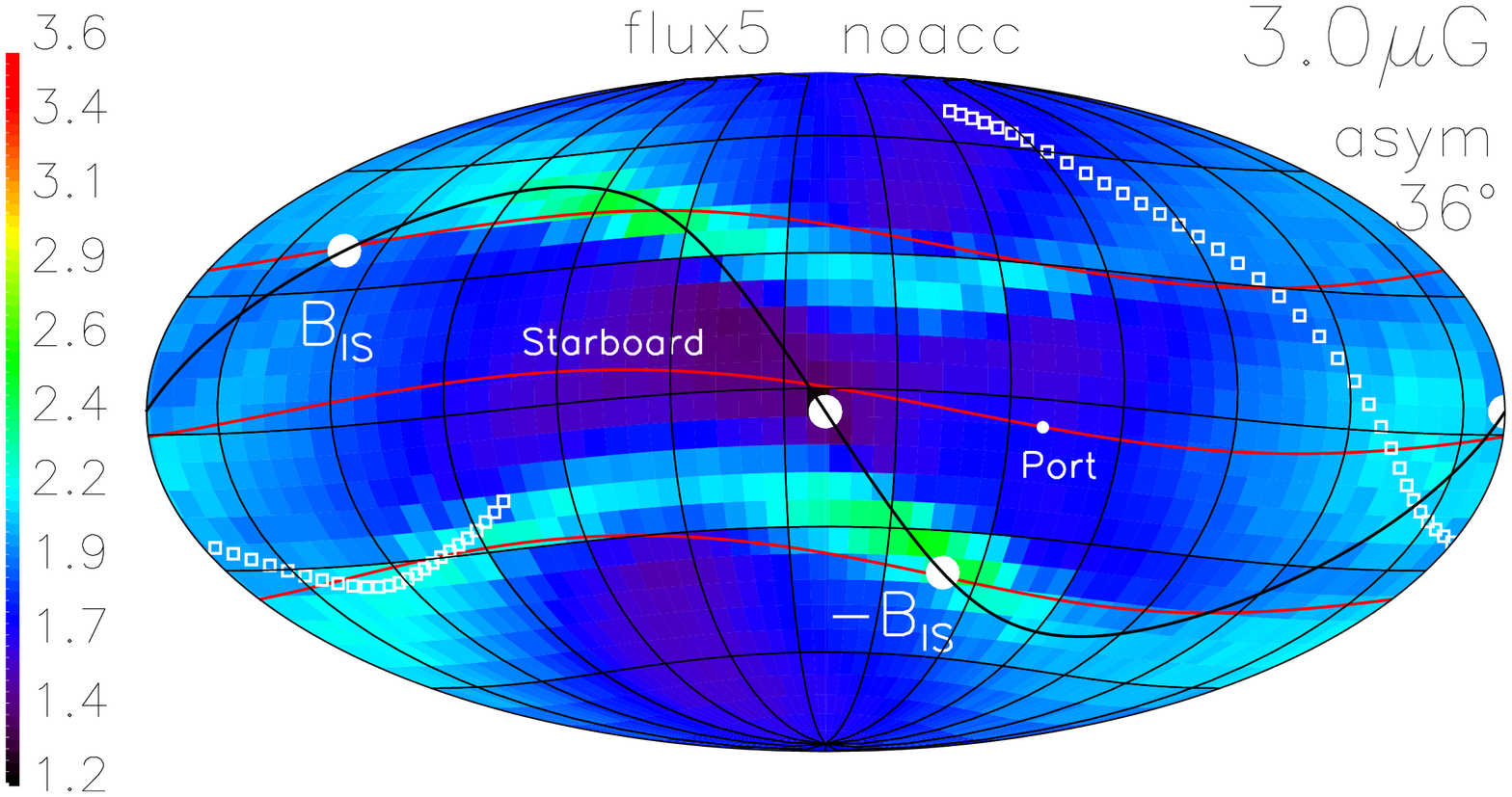}
\caption{ENA flux distribution (4.3 keV) 
in units (cm$^2$ s sr keV)$^{-1}$ for $B_{IS}$
5 and 3 $\mu$G.
The solar wind is asymmetric (fast/slow) with the slow 
wind contained within $\pm$36$^o$ from the equator.
The thin red lines are the projections
of the solar equator and of the boundaries between 
the slow and fast solar wind.
The second figure includes the outline of the heliopause and
of the regions filled by the shocked fast solar wind plasma
at the distance 250 AU from the Sun.
The last two figures show the results ($B_{IS}$=
5 and 3 $\mu$G) with adiabatic acceleration
switched off. 
The values of $V_{IS}$=23.2 km/s, $n_{IS}$=0.06 cm$^{-3}$, 
$n_H$=0.1 cm$^{-3}$, $V_{SW}$=750/400 km/s (fast/slow), 
$n_{SW, 1AU}$=1.58/5.55cm$^{-3}$ (fast/slow) are the same for all figures.
\label{fenaasym}}
%\label{fenaasym}
\end{figure}

The first three years of IBEX data refer to the extended solar
minimum period, when the solar wind asymmetry (two-component
slow + fast structure) was prominent. The results of our calculations
for the case of asymmetric solar wind are collected in 
Figures \ref{fenaasym15}, \ref{fenaasym} and \ref{fenan2}.

For very strong field ($B_{IS}$=15 $\mu$G, Fig. \ref{fenaasym15}), the
peaks from the two plasma streams are prominent.

For $B_{IS}$=5 $\mu$G and less, the directional distribution of the ENA flux 
following from our models 
reflects primarily the two-component solar wind structure. Upstream from the
shock, the slow wind is in our models restricted to the region between
$\pm$36$^o$ solar latitude (red lines in Figs. \ref{fenan2} and 
\ref{fenaasym}). Figure \ref{fenan2} shows that, at low ENA energy (1.1 
keV) this region corresponds to high ENA intensity belt. 
At high ENA energy (4.3 keV. Fig. \ref{fenaasym}) there is a similar 
belt, but of low ENA intensity. 

This structure can be understood by observing that the ENA coming from
directions within the $\pm$36$^o$ solar latitude derive mostly from the 
energetic protons carried by the shocked slow solar wind plasma. The
density of $\sim$1 keV energetic protons is much higher in the shocked
slow wind than in the shocked fast wind (Fig. \ref{fspecsh}). This 
explains the high ENA intensity belt at low ($\sim$1 keV) energy. 
On the other hand, the energetic proton 
flux at $\sim$4.3 keV is approximately the same for the slow and fast 
solar wind plasma, because the high average energy of the energetic 
protons in the fast wind is compensated by the high density of the slow 
wind (Fig. \ref{shpui}). The reason why the 4.3 keV ENA flux is higher 
in the region outside the slow wind belt (Fig. \ref{fenaasym})
is that the convection time
from the termination shock (and consequently the neutralization loss) 
for the energetic protons in the fast wind plasma is lower than in the 
slow wind.

For $B_{IS}$ 2-5 $\mu$G (Fig. \ref{fenaasym}), we find that the low 
intensity belt for 4.3 keV ENA narrows down near the ISM downwind 
direction. This structure is very much alike to the "split tail" 
observed by IBEX (\citet{mccomas2013}, \citet{schwadron2014}).  The low 
ENA flux 
region around the ISM downwind direction is almost split into two parts
(the "port" and the "starboard" lobes: \citet{mccomas2013}). 
We find that this effect is caused by the streams of fast wind plasma 
entering the heliotail region. Figure \ref{fenaasym} (second panel) 
shows the outline of the heliopause (thick black oval) at the distance 250 AU 
from the Sun, together with the outlines of the two regions in the 
heliotail filled by the plasma originating from the fast solar wind (thin
black ovals). The flux of $\sim$4 keV ENA from these fast wind plasma streams 
is higher than from the slow wind plasma region between them. Since the 
fast wind streams at large distances move to lower heliolatitudes, the 
two high ENA flux regions associated with them intrude into the
$\pm$36$^o$ heliolatitude belt and partly split 
the low ENA flux region.

The low 4.3 keV ENA flux region obtained from our calculations (Fig. 
\ref{fenaasym}) is tilted relative to ecliptic. This tilt is not 
dependent on $B_{IS}$ value (see Fig. \ref{fenaasym}). On the other 
hand, it is close to the 7.25$^o$ tilt of the solar equator 
(and of the slow/fast 
solar wind boundary: see the red lines in the figures). A similar tilt 
appears in the IBEX data (\citet{mccomas2013}, Fig. 6). 
It may be concluded that the main cause of the "split tail" structure is 
the distribution of the slow and fast solar wind streams.

Our results concerning the "split tail" structure remain valid when
the effect
of adiabatic acceleration is switched off (Fig. \ref{fenaasym}, two lower
panels). This is important, because we expect that adiabatic 
acceleration is overestimated by our models. 
On the other hand, adiabatic acceleration of the energetic 
protons affects the production of ENA from the region near the apex of 
the ISM (the nose region: Fig. \ref{fenaasym}, two upper panels). 
   
\subsection{Comparison with IBEX}

In addition to the "split tail" structure, our results show other
points of qualitative similarity with IBEX. At low energy, the
ENA flux in our calculations have a peak near the tail direction
(Fig. \ref{fenan2}) which, at higher energy, evolves into a
depletion (Fig. \ref{fenaasym}). A similar behaviour can be seen
in the IBEX data (\citet{schwadron2014}, Fig. 10). 

Compared to the ENA fluxes observed by IBEX, our calculations 
produce lower ENA intensity.
The magnitude of the ENA flux at 4.3 keV obtained in our models
is lower than the non-ribbon flux observed by IBEX by about a factor
of 2 (Fig. 6 in \citet{mccomas2013}, Fig. 10 in \citet{schwadron2014}). 
For $\sim$1.1 keV this factor is about 4.

In Figure \ref{fgam5} we show our results for the slope of the
ENA flux spectrum for the case of 5 $\mu$G 
field, asymmetric solar wind. We plot the values of $\gamma$,
defined as $\gamma=-\log(J_{ENA}(E_1)/J_{ENA}(E_2))/\log(E_1/E_2)$
for ($E_1$,$E_2$)=(1.1 keV,1.7 keV) and (2.7 keV, 4.3 keV). 
The values of the spectral index obtained in our model are lower than 
observed by IBEX 
(\citet{mccomas2013}, \citet{schwadron2014}), 
suggesting that our model underestimates the ENA flux at low energy compared to the
high energy flux. On the other hand, we find that the high $\gamma$ region
is similar to the "split tail" structure in 
the ENA flux distribution,
in agreement with IBEX observations 
(\citet{mccomas2013}, Fig.5; \citet{schwadron2014}, 
Fig. 11, top panel).

\section{Summary and conclusions}

We present a set of 3D time-stationary models of the heliosphere
based on numerical MHD solutions 
for a wide range of interstellar magnetic field strength
(2-20 $\mu$G). 

For the ISMF of 20 $\mu$G and the star at rest, our result is
analogous to a well known Parker solution, with a spherical termination
shock, and the astrosphere elongated along the $\vec{B}_{IS}$ direction.

For strong interstellar field and $V_{IS}\neq0$, we show that the plasma 
flow inside the heliosphere is concentrated in two streams directed close 
(though not exactly 
parallel or antiparallel) to the interstellar field direction.
As a result, the forward part of the heliopause forms straight "wings".
A similar shape was predicted using the Newtonian approximation.

When the field strength decreases to realistic values (2-5 $\mu$G), 
the two-stream structure becomes less prominent. This structure depends 
on the interstellar neutral hydrogen background:
the weaker $n_H$, the more prominent the streams. 

For the simulated heliospheres we calculated the distribution of the 
energetic protons (originating from the pick-up protons from the solar 
wind) and the energetic neutral atom fluxes produced by neutralization 
of these protons. We assumed a simple model of the energetic proton 
distribution at the termination shock. The results were compared with 
the ENA observations by IBEX. Our simulations are restricted to the 
ENA created inside the heliosphere, so that the "ribbon" contribution
must be excluded from the IBEX data. 

We find that the structure of the ENA flux distribution at higher IBEX
energy (1.7 keV and higher) in the 
downwind hemisphere is similar to the "split tail" structure observed
by IBEX (two low ENA flux "lobes" shifted relative to the downwind direction). 
This result persists for different values (2-5 $\mu$G)
of the interstellar magnetic field strength. The explanation suggested
by our model is that this effect follows from the two component
(fast + slow) solar wind structure, prevalent near the low activity
parts of the solar cycle. The tilt of the observed ENA flux structure
is explained by the tilt of the slow/fast solar wind boundary relative to
the ecliptic plane. 

For spherically symmetric solar wind, as observed near solar maxima, the 
ENA distribution in our time-stationary model has a different form. In 
reality, because of long time delay (average $\sim$3 years) before 
neutralization of the parent proton, we expect that the ENA distribution
near solar maximum would still show some effects of the two-component 
solar wind. 

For very strong $B_{IS}$, the ENA distributions have two prominent peaks
corresponding to two plasma streams. 

There are also peaks of the ENA flux that appear in the regions near 
the nose of the heliosphere. However, these effects disappear when 
adiabatic acceleration of energetic protons is neglected. Adiabatic 
acceleration depends on details of plasma flow and density 
distribution and is sensitive to numerical effects.  
The "split tail" structure and the two 
stream-related ENA peaks are not dependent on adiabatic acceleration.

Finally, we note that the recent time-dependent simulation 
\citep{zirnstein2015} found the ENA flux distribution from the heliosphere
with little periodic change over the solar cycle. Their approach was 
different from ours. In particular, the pick-up protons were not treated
separately, but assumed to form a fixed fraction of the bulk plasma.

%% Included in this acknowledgments section are examples of the
%% AASTeX hypertext markup commands. Use \url without the optional [HREF]
%% argument when you want to print the url directly in the text. Otherwise,
%% use either \url or \anchor, with the HREF as the first argument and the
%% text to be printed in the second.

\acknowledgments

We thank M. Strumik for many helpful discussions.
Work in the U.S. was supported by the IBEX mission as part of NASA's
Explorer Program. A.C. was supported by the grant 2012/06/M/ST9/00455
from the Polish National Science Center.


%% in parentheses, after the keyword, but they will not be verified.



\appendix

\section{Energetic ion distribution}

The method is similar to that used in Czechowski et al. 
(\cite{czechowski2012}). The velocity distribution of energetic protons
$f(\vec{r},v)$ is assumed to be isotropic in plasma frame. It 
is obtained by solving the time-stationary transport equation
\begin{equation}
\label{eqtr}
[-\vec{V}\cdot\nabla-\beta_{cx}
+\frac{1}{3}(\nabla\cdot\vec{V})\frac{\partial}{\partial\log v}]f=0
\end{equation}
where $\vec{V}$ is the plasma velocity, $v$ the particle velocity
and $\beta_{cx}$ is the rate for
charge exchange between the energetic protons and low energy neutral 
hydrogen. The equation includes the effects of convection, 
adiabatic energy changes and neutralization loss. We assume that
the proton energy is low enough to neglect spatial diffusion. 

Equation \ref{eqtr} together with the boundary condition at the
termination shock determine the $f(\vec{r},v)$ along a plasma 
streamline $\vec{r}=\vec{r}(s)$ parametrized by the length $s$:
\begin{equation} 
f(\vec{r}(s),v(s))=f_{\rm shock}(\vec{r}_0,v_0)
\exp \left[-\int_0^s{\rm d}s'\frac{\beta_{cx}(s')}{V(\vec{r}(s'))}\right]
\label{eqf} 
\end{equation} 
where $\vec{r}(s)$ and $v(s)$ satisfy the equations 
$V{\rm d}\vec{r}/{\rm d}s=\vec{V}(s)$ and 
$V {\rm d}\log v/{\rm d}s=-(1/3)\nabla\cdot\vec{V}$, with the initial
conditions at the termination shock $\vec{r}(0)$=$\vec{r}_0$, $v(0)$=$v_0$, 
respectively. If the plasma mass flow is 
conserved, the equation for $v(s)$ implies $v(s)\rho(s)^{-1/3}$=const.
Since determination of $\nabla\cdot\vec{V}$ from numerical output is 
imprecise, we use the above relation to determine $v(s)$. 
 
We use a simple analytical model for the energetic proton distribution
at the termination shock $f_{\rm shock}(\vec{r},v)$ in the plasma frame
\begin{equation}
f_{\rm shock}(\vec{r}, v)=n_{\rm PUI}(\vec{r})F_\kappa(v,w)
+n_{\rm SW}(\vec{r})F_G(v,v_T)
\end{equation}
The first term is the distribution of the accelerated pick-up protons
(number density $n_{PUI}$), and the second describes the protons from the
shock-heated bulk solar wind (number density $n_{SW}$).
$F_\kappa(v,w)$ is the kappa function 
\begin{equation}
F_\kappa(v,w)=\frac{\Gamma(\kappa+1)}{\Gamma(\kappa-1/2)}
\frac{1}{(\pi\kappa w^2)^{3/2}}
\left(1+\frac{v^2}{\kappa w^2}\right)^{-(\kappa+1)}
\end{equation}
We choose $\kappa$=1.65. The "thermal speed" parameter {\bf $w$} we 
determine from the requirement that the total pick-up proton energy 
should be equal to 0.8 of the energy of the solar wind upstream from the 
shock, as observed by
Voyager 2 \citep{richardson2008}.
The bulk solar wind protons are described by the Maxwellian
distribution $F_G(v,v_T)=\exp -(v/v_T)^2/\pi^{3/2}v_T^3$ with the thermal 
speed $v_T$ equal to one half of the solar wind
speed downstream. The number density $n_{SW}$ is taken 
from the MHD model.

\section{ENA flux estimations}

Assume that the plasma flow is incompressible, so
that Eq. \ref{eqtr}  becomes
\begin{equation}
\vec{V}\cdot\nabla f=-\beta_{cx}f
\end{equation}
The same equation holds for the ion flux $J_{ion}$.

Consider first the ENA flux from the direction towards the nose of 
the heliosphere, along the stagnation line.
Let the plasma velocity decrease linearly towards
the heliopause: $V=V_0(1-z/L)$ where $z$ is the distance from the shock
along the stagnation line
and $L$ the distance from the shock to the heliopause. 

The equation for the ion flux is
\begin{equation}
\frac{d J_{ion}}{dz}=\frac{\beta_{cx}}{V_0(1-z/L)}J_{ion}
\end{equation}
with the solution
\begin{equation}
J_{ion}(z)=J_0 \left(1-z/L\right)^{\beta_{cx}L/V_0}
\end{equation}
The ENA flux is then given by (see Eq. \ref{eqjena})
\begin{equation}
\label{enanose}
J_{ENA}=\frac{\beta_{cx}L/V_0}{1+\beta_{cx}L/V_0}\frac{V_0}{v}J_0
\end{equation}
where $v$ is the particle speed and we have used 
$\beta_{cx}=\sigma_{cx} v n_H$.
 
Consider next the line-of-sight within one of the plasma streams
assuming that the plasma speed $V$ is constant and parallel to the 
line-of-sight. The solution for the ion flux is then
\begin{equation}
J_{ion}(z)=J_0 \exp (-z\beta_{cx}/V)
\end{equation}
and for the ENA flux
\begin{equation}
\label{enastr}
J_{ENA}=\frac{V_0}{v}J_0
\left(1-\exp \left(-\frac{\beta_{cx}L}{V}\right)\right)
\end{equation}
When the distance to the heliopause along the line-of-sight is 
large compared to $V/\beta_{cx}$, the ENA flux becomes 
$J_{ENA}=(V/v)J_0$.




\clearpage

\begin{figure}
\centering
%\epsscale=0.4
%\plotone{figs/sena3asym_noa.eps}
\includegraphics[width=8cm]{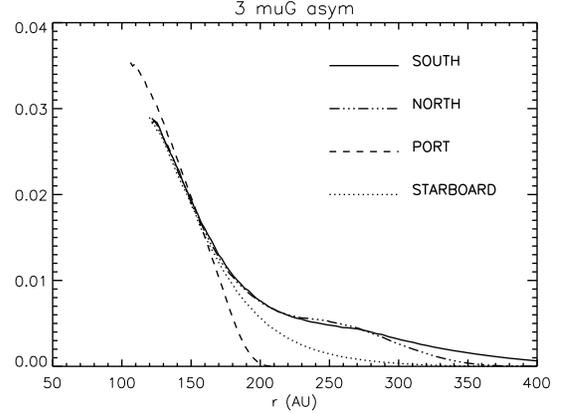}
\caption{Profiles of the ENA production rate (4.3 keV)
along 4 different directions corresponding to: the port
lobe; the starboard lobe; the "pinch" from the north, and the
"pinch" from the south. The case illustrated 
corresponds to the lowest panel in Fig. \ref{fenaasym}
($B_{IS}$=3 $\mu$G, asymmetric solar wind, adiabatic acceleration
switched off).   
\label{fpinch}}
%\label{fpinch}
\end{figure}

\begin{figure}
\centering
%\epsscale=0.4
%\plotone{figs/gam3_bis5asym.eps}
%\plotone{figs/gam5_bis5asym.eps}
\includegraphics[width=7cm]{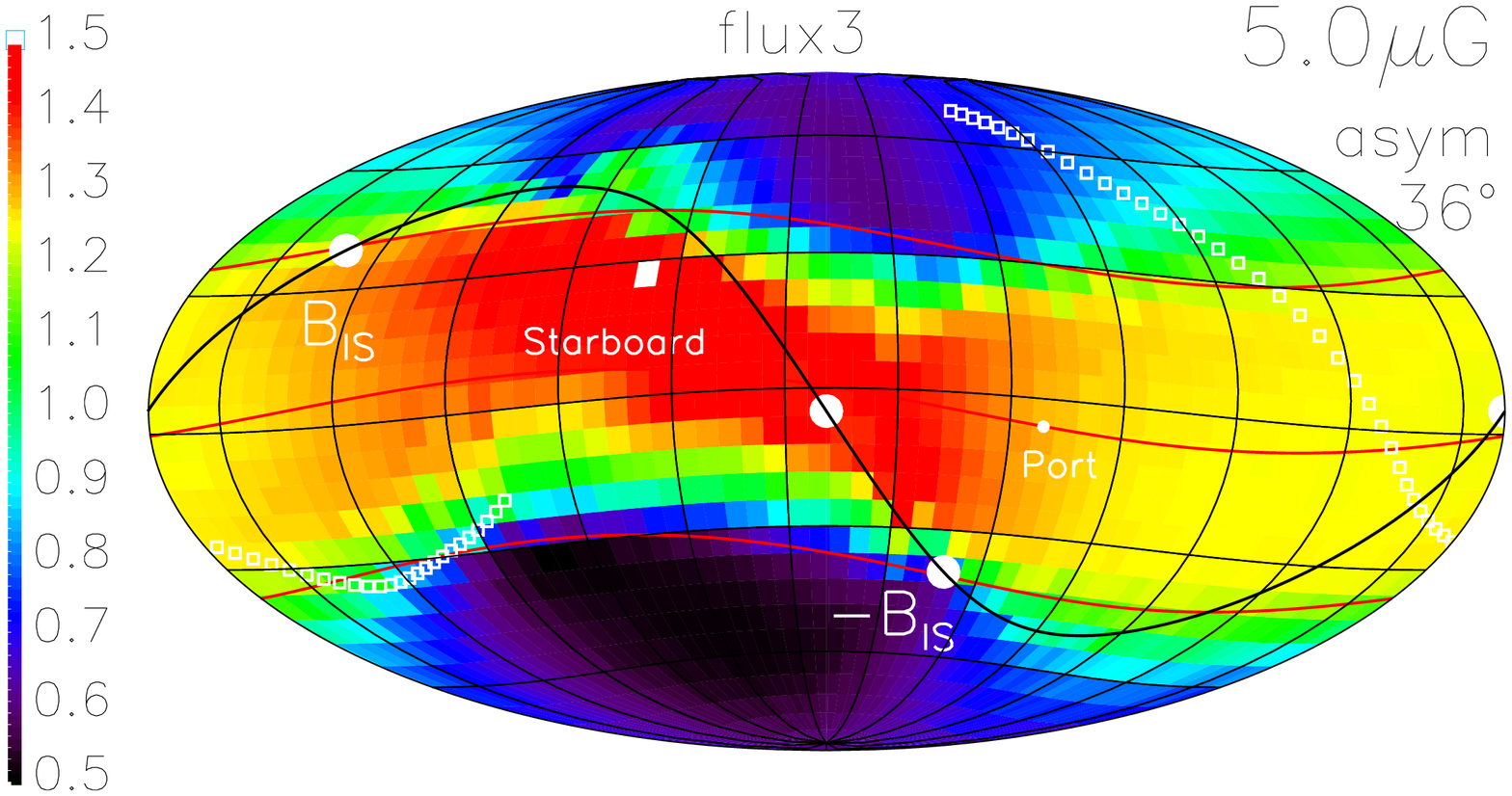}
\includegraphics[width=7cm]{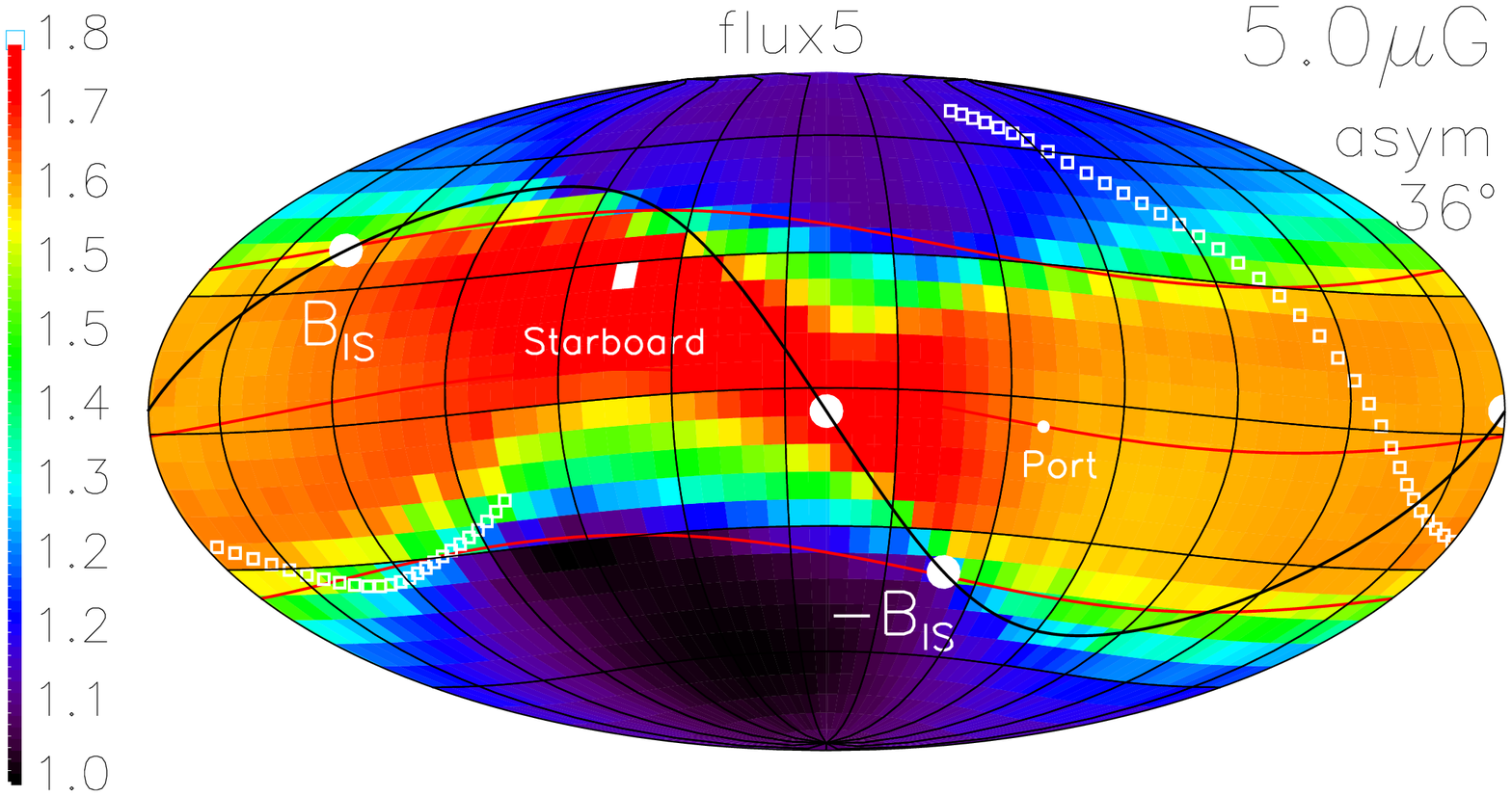}
\caption{Spectral index distributions in the 1.1-1.7 keV (first panel)
and in the 2.7-4.3 keV (second panel) energy ranges 
for numerical solution 
with $B_{IS}$=5 $\mu$G, asymmetric (fast+slow) solar wind, 
$n_H$=0.1 cm$^{-3}$, $V_{IS}$=23.2 km/s, $n_{IS}$=0.06 cm$^{-3}$, 
$V_{SW}$=750/400 km/s (fast/slow), 
$n_{SW, 1AU}$=1.58/5.55cm$^{-3}$ (fast/slow).
The projection is the same as in Figs. 5 and 6 of 
\citep{mccomas2013}). 
\label{fgam5}}
%\label{fgam5}
\end{figure}

\clearpage

%% If you use the table environment, please indicate horizontal rules using
%% Do not put multiple tabular environments within a single table.
%% The optional \label should appear inside the \caption command.

\begin{table}
\caption{Parameters and results of MHD calculations\label{table1}}
%\label{table1}
\begin{center}          
%\begin{tabular}{l c c c c c c c c c c} 
\begin{tabular}{l l l l l l l l l l l} 
\tableline\tableline          
$B_{IS}$ & $n_{IS}$ & $n_H$ & $V_{IS}$ & $V_{SW}$ & $n_{SW, 1 AU}$
& r$_{\rm TS,min}$ & r$_{\rm TS,max}$ & r$_{\rm HP,min}$ & height & width\\ 
($\mu$G) & (cm$^{-3}$) & cm$^{-3}$ & (km/s) & (km/s) & (cm$^{-3}$) 
& (AU) & (AU) & (AU) & (AU) & (AU)\\ 
& & & & fast/slow & fast/slow & & & & &\\ 
\tableline                        
   20 & 0.04 & 0.1 & 0 & 750 & 4.2 & 36 & 40 & 95 & &\\   
   20 & 0.06 & 0.1 & 23.2 & 750 & 4.2 & 36 & 43 & 57 & 140 & 749\\   
   20 & 0.06 & 0.01 & 23.2 & 750 & 4.2 & 35 & 41 & 65 & 154 & 1840\\   
\tableline
   15 & 0.06 & 0.1 & 23.2 & 400 & 5.55 & 31 & 38 & 47 & 113 & 444\\
   15 & 0.06 & 0.1 & 23.2 & 750/400 & 1.58/5.55 & 31 & 39 & 45 & 130 & 483\\
\tableline
   5 & 0.06 & 0.1 & 23.2 & 400 & 5.55 & 63 & 91 & 97 & 282 & 622\\
   5 & 0.06 & 0.001 & 23.2 & 400 & 5.55 & 75 & 109 & 109 & 317 & 1827\\
   5 & 0.06 & 0.1 & 23.2 & 750/400 & 1.58/5.55 & 62 & 92 & 93 & 315 & 693\\
\tableline
   3 & 0.06 & 0.1 & 23.2 & 400 & 5.55 & 75 & 122 & 115 & 384 & 643\\
   3 & 0.06 & 0.1 & 23.2 & 750/400 & 1.58/5.55 & 74 & 120 & 111 & 424 & 719\\
\tableline
   2 & 0.06 & 0.1 & 23.2 & 400 & 5.55 & 81 & 140 & 119 & 469 & 654 \\
   2 & 0.06 & 0.1 & 23.2 & 750/400 & 1.58/5.55 & 76 & 135 & 111 & 536 & 720 \\
\tableline                                   
\end{tabular}
\end{center}
\end{table}

%% If the table is more than one page long, the width of the table can vary
%% from page to page when the default \tablewidth is used, as below.  The
%% individual table widths for each page will be written to the log file; a
%% maximum tablewidth for the table can be computed from these values.
%% The \tablewidth argument can then be reset and the file reprocessed, so
%% that the table is of uniform width throughout. Try getting the widths
%% from the log file and changing the \tablewidth parameter to see how
%% adjusting this value affects table formatting.

%% The \dataset{} macro has also been applied to a few of the objects to
%% show how many observations can be tagged in a table.

%% Tables may also be prepared as separate files. See the accompanying
%% sample file table.tex for an example of an external table file.
%% To include an external file in your main document, use the \input
%% command. Uncomment the line below to include table.tex in this
%% sample file. (Note that you will need to comment out the \documentclass,
%% \begin{document}, and \end{document} commands from table.tex if you want
%% to include it in this document.)


%% The following command ends your manuscript. LaTeX will ignore any text


\begin{thebibliography}{}
\bibitem[Burlaga \& Ness(2014)]{burlaga_ness2014} Burlaga, L.F., Ness, N.F., 
%2014, \apj, 784, 146
%Voyager 1 observations of the interstellar magnetic field and the transition from
%the heliosheath,
2014, \apj, 784, 146
\bibitem[Bzowski(2008)]{bzowski2008}
Bzowski, M.,
%Survival probablility and energy modification of hydrogen energetic neutral atoms
%on their way from the termination shock to Earth orbit,
2008, Astronomy \& Astrophysics, 488, 1057
\bibitem[Bzowski et al.(2009)]{bzowski2009}
Bzowski, M., M\"obius, E., Tarnopolski, S., Izmodenov, V., Gloeckler, G.,
%Neutral H density at the termination shock: A consolidation of recent results,
2009, Space Science Revievs, 143, 177
\bibitem[Bzowski et al.(2012)]{bzowski2012}
Bzowski, M., Kubiak, M.A., M\"obius, E., Bochsler, P., Leonard, T., Heirtzler, D.,
Kucharek, H., Sokol, J.M., Hlond, M., Crew, G.B., Schwadron, N.A., Fuselier, S.A.,
McComas, D.J.,
%Neutral interstellar helium parameters based on IBEX-Lo observations and test
%particle calculations,
2012, \apjs, 198, 12
\bibitem[Czechowski \& Grzedzielski(1998)]{czechowski1998}
Czechowski, A., Grzedzielski, S.,
%Can the direction of interstellar magnetic field be determined from the
%heliotail ENA flux?,
1998, \grl, 25, 1855
\bibitem[Czechowski, Hilchenbach \& Hsieh (2012)]{czechowski2012}
Czechowski, A., Hilchenbach, M., Hsieh, K.C.,
%HSTOF ENA observations and energetic ion distributions in the heliosheath,
2012, Astronomy \&  Astrophysics, 541, 14
\bibitem[Dell, Foley \& Ruderman(1965)]{drell1965}
Drell, S.D., Foley, H.M., Ruderman, M.A.,
%Drag and propulsion of large satellites in the ionoosphere; An Alfv\'en
%propulsion engine in space,
1965, \jgr, 70, 3131
\bibitem[Fahr, Grzedzielski \& Ratkiewicz-Landowska(1988)]{fahr1988}
Fahr, H.J., Grzedzielski, S., Ratkiewicz-Landowska, R.,
%Magnetohydrodynamic modeling of the 3-dimensional heliopause using the 
%Newtonian approximation,
1988, Ann. Geophys., 6, 337
\bibitem[Florinski et al.(2004)]{florinski2004}
Florinski, V., Pogorelov, N.V., Zank, G.P., Wood, B.E., Cox, D.P.,
%On the possibility of a strong magnetic field in the local interstellar medium,
2004, \apj, 604, 700
\bibitem[Funsten et al.(2009)]{funsten2009} Funsten, H. O., Allegrini, F., Crew, G. B., 
DeMajistre, R., Frisch, P. C., et al.,
2009, Science, 326, 964
\bibitem[Gruntman et al.(2001)]{gruntman2001}
Gruntman, M., Roelof, E.C., Mitc hell, D.G., Fahr, H.J., Funsten, H.O., 
McComas, D.J.,
%Energetic neutral atom imaging of the heliospheric boundary region,
2001, \jgr, 106, 15767
\bibitem[Grygorczuk, Czechowski \& Grzedzielski(2014)]{grygorczuk2014}
Grygorczuk, J., Czechowski, A., Grzedzielski, S.,
%Why are the magnetic field directions measured by Voyager 1 on both
%sides of the heliopause so similar?,
2014, \apjl, 789, L43 
%\bibitem[2011]{grygorczuk2011} Grygorczuk, J., Ratkiewicz, 
%R., Strumik, M., and Grzedzielski, S. 
%2011, \apj, 727:L48,
\bibitem[heerikhuisen et al.(2010)]{heerikhuisen2010}
Heerikhuisen, J., Pogorelov, N., Zank, G.P., Crew, G.B., Frisch, P.C., 
Funsten, H.O, Janzen, P.H., McComas, D.J., Reisenfeld, D.B., Schwadron, N.A.,
%Pick-up ions in the outer heliosheath: A possible mechanism for the Interstellar
%Boundary EXplorer Ribbon,
2010, \apjl, 708, L126 
\bibitem[Heerikhuisen et al.(2014)]{heerikhuisen2014}
Heerikhuisen, J., Zirnstein, E.J., Funsten, H.O, Pogorelov, N.V., Zank, G.P.,
%The effect of new interstellar medium parameters on the heliosphere and energetic
%neutral atoms from the interstellar boundary,
2014, \apj, 784:73, doi:10.1088/004-637X/784/1/73 
\bibitem[Izmodenov(2009)]{izmodenov2009}
Izmodenov, V.V.,
%Local interstellar parameters as they are inferred from analysis of observations 
%inside the heliosphere,
2009, Space Sci. Revs., 143, 139 
\bibitem[Kivelson \& Jia(2013)]{kivelson2013}
Kivelson, M.G., Jia, X.,
%An MHD model of Ganymede's mini-magnetosphere suggests that the heliosphere 
%forms in a sub-Alfv\'enic flow,
2013, \jgr, 0, 0(submitted)
\bibitem[Krimigis et al.(2009)]{krimigis2009}
Krimigis, S.M., Mitchell, D.G., Roelof, E.C., Hsieh, K.C., McComas, D.J.,
%Imaging the interaction of the heliosphere with the interstellar medium from 
%Saturn with Cassini,
2009, Science, 326, 971
\bibitem[Le Chat, Issautier \& Meyer-Vernet(2012)]{lechat2012}
Le Chat, G., Issautier, K., Meyer-Vernet, N.,
%The solar wind energy flux,
2012, Solar Phys., 279, 197
\bibitem[McComas et al.(2009)]{mccomas2009}
McComas, D.J., F. Allegrini, P. Bochsler, M. Bzowski, E.R. Christian, G.B. Crew,
R. DeMajistre, H. Fahr, H. Fichtner, P.C. Frisch, H.O. Funsten, S.A. Fuselier,
G. Gloeckler, M. Gruntman, J. Heerikhuisen, V. Izmodenov, P. Janzen, P. Knappenberger,
S. Krimigis, H. Kucharek, M. Lee, G. Livadiotis, S. Livi, R.J. MacDowall, D. Mitchell,
E. M\"obius, T. Moore, N.V. Pogorelov, D. Reisenfeld, E. Roelof, L. Saul, N.A. Schwadron,
P.W. Valek, R. Vanderspek, P. Wurz, G.P. Zank,
%Global observations of the interstellar interaction from the Interstellar 
%Boundary Explorer (IBEX),
2009, Science, 326, 959
\bibitem[McComas et al.(2012)]{mccomas2012a}
McComas, D.J., Alexashov, D., Bzowski, M., Fahr, H., Heerikhuisen, J., 
Izmodenov, V., Lee, M.A., M\"obius, E., Pogorelov, N., Schwadron, N.A.,
Zank, G.P.,
%The heliosphere's interstellar interaction: No bow shock,
2012, Science, 336, 1291 
\bibitem[McComas et al.(2012b)]{mccomas2012b}
McComas, D.J., Dayeh, M.A., Allegrini, F., Bzowski, M., DeMajistre, R., 
Fujiki, K., Funsten, H.O., Fuselier, S.A., Gruntman, M., Janzen, P.H.,
Kubiak, M.A., Kucharek, H., Livadiotis, G., M\"obius, E., Reisenfeld, D.B.,
Reno, M., Schwadron, N.A., Sokol, J.M., Tokumaru, M.,
%The first three years of IBEX observations and our evolving heliosphere, 
2012, \apjs, 203, 1
\bibitem[McComas et al.(2013)]{mccomas2013}
McComas, D.J., Dayeh, M.A., Funsten, H.O., Livadiotis, G., Schwadron, N.A.,
%The heliotail revealed by the Interstellar Boundary Explorer,
2013,\apj, 771, 77
\bibitem[M\"obius et al.(2012)]{moebius2012}
M\"obius, E., Bochsler, P., Bzowski, M., Heirtzler, D., Kubiak, M.A., Kucharek, H.,
Leonard, T., Lee, M.A., Leonard, T., Schwadron, N.A., Wu, X., Fuselier, S.A., 
Crew, G., McComas, D.J., Petersen, L., Saul, L., Valovcin, D., Vanderspek, R., 
Wurz, P.,
%Interstellar gas flow parameters derived from Interstellar Boundary Explorer-Lo
%observations in 2009 and 2010: analytical analysis,
2012, \apjs, 198, 11
\bibitem[Neubauer(1980)]{neubauer1980}
Neubauer, F.M.,
%Nonlinear standing Alfv\'en wave current system at Io: Theory,
1980, \jgr, 85, 1171
\bibitem[Parker(1961)]{parker61}
Parker, E.N., 
%The stellar wind regions, 
1961, \apj, 134, 20
\bibitem[Pogorelov et al.(2011)]{pogorelov2011}
Pogorelov, N.V., heerikhuisen, J., Zank, G.P., Borovikov, S.N., Frisch, P.C.,
McComas, D.J.,
%Interstellar boundary expolrer measurements and magnetic field in the vicinity 
%of the heliopause,
2011, \apj, 742, 104, doi: 10.1088/0004-637X/742/2/104
\bibitem[Richardson et al.(2008)]{richardson2008}
Richardson, J.D., Kasper, J.C., Belcher, J.W., Lazarus, A.J.,
%Cool heliosheath plasma and deceleration of the upstream solar wind at the 
%termination shock,
2008, Nature, 454, 63-66
\bibitem[Ratkiewicz et al.(1998)]{ratkiewicz98} Ratkiewicz, R., Barnes, A., 
Molvik, G. A.,Spreiter, J. R., Stahara, S. S.,
Vinokur, M., and Venkatesvaran, S. A.,
1998, A\&A, 335, 363-369
\bibitem[Saur et al.(2013)]{saur2013}
Saur, J., Grambusch, T., Duling, S., Neubauer, F.M., Simon, S.,
%Magnetic energy fluxes in sub-Alfvenic planet star and moon planet 
%interactions,
2013, Astr. Astrophys., 552, A119
\bibitem[Schwadron et al.(2009)]{schwadron2009}
Schwadron, N.A., Bzowski, M., Crew, G.B., Gruntman, M., Fahr, H., Fichtner, H.,
P.C. Frisch, H.O. Funsten, S. Fuselier, J. Heerikhuisen, V. Izmodenov, H. Kucharek,
M. Lee, G. Livadiotis, McComas, D.J., E. M\"obius, T. Moore, J. Mukherjee,
N.V. Pogorelov, C. Prested, D. Reisenfeld, E. Roelof, G.P. Zank, 
%Comparison of Interstellar Boundary Explorer observations with 3D global 
%heliospheric models, 
2009, Science, 326, 966
\bibitem[Schwadron et al.(2011)]{schwadron2011}
Schwadron, N.A., Allegrini, F., Bzowski, M., Christian, E.R., Crew, G.B., Dayeh, M.,
DeMajistre, R., Frisch, P., Funsten, H.O., Fuselier, S.A.,
Goodrich, K., Gruntman, M., Janzen, P., Kucharek, H., Livadiotis, G., McComas, D.J., 
M\"obius, E., Prested, C., Reisenfeld, D., Reno, M., Roelof, E., Siegel, J., 
Vanderspek, R.,
%Separation of the Interstellar Boundary Explorer Ribbon from Globally Distributed 
%Neutral Atom Flux,
2011, \apj, 731, 56
\bibitem[Schwadron et al.(2014)]{schwadron2014}
Schwadron, N.A., M\"obius, E., Fuselier, S.A., McComas, D.J., Funsten, H.O.,
Janzen, P., Reisenfeld, D., Kucharek, H., Lee, M.A., Fairchild, K., 
Allegrini, F., Dayeh, M., Livadiotis, G., Reno, M.,
Bzowski, M., Sokol, J.M., Kubiak, M.A., Christian, E.R., 
DeMajistre, R., Frisch, P., Galli, A., Wurz, P., Gruntman, M.,    
%Separation of the Ribbon from Globally Distributed Energetic 
%Neutral Atom Flux using the first five years of IBEX observations,
2014, \apjs, 215, 13
\bibitem[Stone et al.(2008)]{stone2008}
Stone, E.C., Cummings, A.C., McDonald, F.B., Heikkila, B.C., Lal, N., 
Webber, W.R.,
%An asymmetric solar wind termination shock
2008, Nature, 454, 71-74
\bibitem[Thomas(1978)]{thomas78}
Thomas, G.E.,
%The interstellar wind and its influence on the interplanetary environment,
1978, Ann. Rev. Earth Planet. Sci. 6, 173-204
\bibitem[Wood et al.(2014)]{wood2014}
Wood, B.E., Izmodenov, V.I., Alexashov, D.B., Redfield, S., Edelman, E.,
%A new detection of Ly$\alpha$ absorption from the heliotail,
2014, \apj, 780, 108
\bibitem[Zank et al.(2013)]{zank2013}
Zank, G.P., Heerikhuisen, J., Wood, B.E., Pogorelov, N.V., Zirnstein, E., 
McComas, D.J.,
%Heliospheric structure: the bow wave and the hydrogen wall,
2013, \apj, 763, 20 
\bibitem[Zirnstein et al.(2015)]{zirnstein2015}
Zirnstein, E.J., Heerikhuisen, J., Pogorelov, N.V., McComas, D.J., Dayeh, M.A.,
%Simulations of a dynamic solar cycle and its effects on the Interstellar Boundary 
%Explorer ribbon and globally distributed energetic neutral atom flux,
2015, \apj, 804, 5

\end{thebibliography}
\end{document}